\documentclass[useAMS,usenatbib]{mn2e}
\usepackage{times}
\usepackage{graphicx}
\usepackage{subfigure}
\usepackage{psfrag}

\title[WMAP 3-year primordial power spectrum]{WMAP 3-year primordial power spectrum}
\author[M. Bridges et al.]
  {M. Bridges,$^1$\thanks{E-mail: m.bridges@mrao.cam.ac.uk}
   A.N. Lasenby,$^1$ M.P. Hobson$^1$\\
  $^1$Astrophysics Group,
      Cavendish Laboratory, Madingley Road,
      Cambridge CB3 0HE, UK\\
}

\date{Accepted ---. Received ---; in original form \today}
\pagerange{\pageref{firstpage}--\pageref{lastpage}}
\pubyear{2006}

\begin{document}
\label{firstpage}
\maketitle

\begin{abstract}

We constrain the form of the primordial power spectrum using Wilkinson Microwave Anisotropy
Probe (WMAP) 3-year cosmic microwave background (CMB) data (+ other high resolution CMB
experiments) in addition to complementary large-scale structure (LSS) data: 2dF, SDSS, Ly-$\alpha$
forest and luminous red galaxy (LRG) data from the SDSS catalogue. We extend the work of the
WMAP team to that of a fully Bayesian approach whereby we compute the
comparative Bayesian evidence in addition to parameter estimates for a collection of seven
models: (i) a scale invariant Harrison-Zel'dovich (H-Z) spectrum; (ii) a power-law; (iii) a
running spectral index; (iv) a broken spectrum; (v) a power-law with an abrupt cutoff on
large-scales; (vi) a reconstruction of the spectrum in eight bins in wavenumber; and (vii) a
spectrum resulting from a cosmological model proposed by \citet{Doran} (L-D). 
Using a basic dataset of WMAP3 + other CMB + 2dF + SDSS our analysis
confirms that a scale-invariant spectrum is disfavoured by between 0.7 and 1.7 units
of log evidence (depending on priors chosen)
when compared with a power-law tilt. Moreover a running spectrum is now significantly
preferred, but only when using the most constraining set of priors. The addition of
Ly-$\alpha$ and LRG data independently both suggest much lower values of the running index
than with basic dataset alone and interestingly the inclusion of Ly-$\alpha$
significantly disfavours a running parameterisation by more than a unit in log evidence.
Overall the highest evidences, over all datasets, were obtained with a power law spectrum
containing a cutoff with a significant log evidence difference of roughly 2 units. The
natural tilt and exponential cutoff present in the L-D spectrum is found to be favoured 
decisively by a log evidence difference of over 5 units, but only for a limited study within the best-fit
concordance cosmology. 

\end{abstract}


\section{Introduction}
The recent release of 3-year Wilkinson Microwave Anisotropy Probe (WMAP3;
\citealt{WMAP3}) data have provided precise measurements of temperature fluctuations
in the cosmic microwave background (CMB). The accepted inflationary paradigm suggests
that a primordial spectrum of almost scale-invariant density fluctuations produced
during inflation went on to produce the observed structure in the CMB and that seen
on large-scales in the current distribution of matter. Now, for the first time a
purely scale invariant primordial spectrum is ruled out at $1\sigma$
(\citealt{SpergelII}; \citealt{Parkinson}) in favour of a `tilted' spectrum with $n <
1$. The WMAP team have already attempted limited constraints on the form of the
spectrum and \citet{Parkinson} have conducted a model selection study to ascertain
the necessity of a tilt in the spectrum with the new data. In this paper we extend
both studies to a suite of models covering a wide variety of possibilities based on
both physical and observational grounds. We use a fully Bayesian approach to
determine the model parameters and comparative evidence to ascertain which model the
data actually prefers.

Our previous paper \citep{Bridges} [Bridges06] used WMAP 1-year data (WMAP1;
\citealt{WMAP1}) to constrain the same set of models.  These generalisations were
motivated principally by observations of a decrement in power on large-scales from
WMAP1 and a tilting spectrum on small-scales from high resolution experiments such as
the Arcminute Cosmology Bolometer Array (ACBAR; \citealt{ACBAR}), the Very Small
Array (VSA ; \citealt{VSA}) and the Cosmic Background Imager (CBI; \citealt{CBI}).
With two more years observing time and improved treatment of systematic errors the
decrement in power on large scales is now somewhat reduced, yet still evident in
WMAP3 and is now constrained almost to the cosmic variance limit while the tilting
spectrum on small-scales is now seen even without the aid of high-resolution small
scale experiments, due to tighter constraints on the second acoustic peak.  On
physical grounds we test a broken spectrum caused perhaps by double field inflation
\citep{barriga} and a spectrum predicted by \citet{Doran} (L-D) naturally
incorporating an exponential cutoff in power on large scales by considering the
evolution of closed universes out of a big bang singularity, with a novel boundary
condition that restricts the total conformal time available in the universe. We also
aim to reconstruct the spectrum in a number of bins in wavenumber $k$.

\section{Model Selection Framework}
Bayesian model selection is now well established within the community as a reliable means of
appropriately determining the most efficient parameterisation for a model, penalising any
unecessary complication (\citealt{Jaffe}, \citealt{Drell}, \citealt{John}, \citealt{Bridle},
\citealt{McLachlan}, \citealt{Slosar}, \citealt{Saini}, \citealt{Marshall},
\citealt{Niarchou}, \citealt{Basset}, \citealt{Mukherjee}, \citealt{Trotta},
\citealt{Beltran}, Bridges06).  Recently much progress has been made in improving the
speed and accuracy of evidence results \citep{Parkinson} by implementing the method of
\citet{Skilling} known as \emph{nested sampling}. In this paper we will employ our own
implementation of this method \citep{Shaw} to evaluate the evidence and use standard Markov
Chain Monte Carlo (MCMC) to make parameter constraints. 

\subsection{Markov Chain Monte Carlo sampling}
\label{MCMCsec}
A Bayesian analysis provides a coherent
approach to estimating the values of the parameters, $\mathbf{\Theta}$, and their
errors and a method for determining which model, $M$, best describes the data, $\mathbf{D}$.
Bayes theorem states that
\begin{equation} P(\mathbf{\Theta}|\mathbf{D}, M) =
\frac{P(\mathbf{D}|\mathbf{\Theta},
M)P(\mathbf{\Theta}|M)}{P(\mathbf{D}|M)},
\end{equation}
where $P(\mathbf{\Theta}|\mathbf{D}, M)$ is the posterior,
$P(\mathbf{D}|\mathbf{\Theta}, M)$ the likelihood,
$P(\mathbf{\Theta}|M)$ the prior, and $P(\mathbf{D}|M)$ the
Bayesian evidence. Conventionally, the result of a Bayesian parameter
estimation is the posterior probability distribution given by the
product of the likelihood and prior. In addition however, the posterior distribution may be used
to evaluate the Bayesian evidence for the model under consideration.

We will employ a MCMC sampling procedure to
explore the posterior distribution using an adapted version
of the {\sc cosmoMC} package \citep{cosmomc} with four CMB
datasets; WMAP3, ACBAR
the VSA and CBI. We also include
the 2dF Galaxy Redshift Survey \citep{2dF}, the Sloan Digital Sky Survey \citep{sloan} and the Hubble Space Telescope (HST) key project
\citep{HST}. These set of experiments comprise dataset I.
Additionally we include two datasets which cover different
scales and probe independent sources.  The Ly-$\alpha$ forest
(\citealt{McDonaldI}; \citealt{McDonaldII})
 (which combined with dataset I makes up dataset II) comprises cosmological absorption by neutral hydrogen observed in
quasar spectra in the intergalactic
medium. It probes fluctuation scales that are small ($\sim$ Mpc) in
comparison to the other datasets used at redshifts between 2-4
so that primordial information has not been erased by
non-linear evolution. It thus provides a very useful
complementary observation when constraining the form of the
primordial spectrum. Previous authors (\citealt{Viel};
\citealt{Seljak}) have already examined this dataset in
conjunction with WMAP3 and others and found that most of the
interesting results observed by \citet{SpergelII}, namely
lowered scalar amplitude and a non-vanishing running index can
be removed when Ly-$\alpha$ is included.  Observations of luminous red galaxies (LRG) 
\citealt{TegmarkII}
(when combined with dataset I becomes dataset III) consists of $>$ 46,000 galaxies taken from the
full SDSS catalogue which represent a highly uniform galaxy
sample containing only luminous early-types over the entire
redshift range studied constituting an excellent tracer of
large scale  structure. \citet{TegmarkII} recently detected
baryonic acoustic oscillations in the
matter power spectrum extracted from this dataset, providing a
welcome confirmation of early universe physics in large scale structure
data. 

In addition to the
primordial spectrum parameters, we parameterise each model using
the following five cosmological parameters; the physical baryon density $\Omega_b
h^2$; the physical cold dark matter density $\Omega_{c} h^2$; the curvature density $\Omega_k$; the Hubble parameter $h$ ($H_0 = h \times100 \mbox{kms}^{-1}$) and the redshift of
re-ionisation $z_{re}$.

\subsection{Bayesian evidence and nested sampling}
\label{bayes}
The Bayesian evidence is the average likelihood over the
entire prior parameter space of the model:
\begin{equation}
\int\int{\mathcal{L}(\Theta_{\rm{C}},\Theta_{\rm{B}})P(\Theta_{\rm{C}})P(\Theta_{\rm{B}})}d^N\Theta_{\rm{C}} d^M\Theta_{\rm{B}},
\label{equation:evidence}
\end{equation}
where $N$ and $M$ are the number of cosmological and Bianchi parameters respectively.
Those models having large areas of prior parameter space with high likelihoods will produce high evidence values and \emph{vice versa}.
 This effectively
penalises models with excessively large parameter spaces, thus naturally incorporating Ockam's razor.

The method of nested sampling is capable of much higher accuracy than previous methods such as thermodynamic
integration (see e.g. \citealt{Beltran}, Bridges06). due to a computationally more
effecient mapping of the integral in Eqn. \ref{equation:evidence} to a single dimension by a
suitable re-parameterisation in terms of the prior \emph{mass} $X$. This mass can be divided
into elements $dX = \pi(\mathbf{\Theta})d^N \mathbf{\Theta}$ which can be combined in any
order to give say
\begin{equation}
X(\lambda) = \int_{\mathcal{L\left(\mathbf{\Theta}\right) > \lambda}} \pi(\mathbf{\Theta}) d^N
\mathbf{\Theta},
\end{equation}
the prior mass covering all likelihoods above the iso-likelihood curve
$\mathcal{L} = \lambda$.  We also require the function
$\mathcal{L}(X)$ to be a singular decreasing function (which is
trivially satisfied for most posteriors) so that using sampled points
we can estimate the evidence via the integral:
\begin{equation}
\mathcal{Z}=\int_0^1{\mathcal{L}(X)}dX.
\label{equation:nested}
\end{equation}
Via this method we can obtain evidences with an accuracy 10 times higher than previous methods 
for the same number of likelihood evaluations.
 
A standard scenario in Bayesian model selection would require the computation
of evidences for two models A and B. The difference of log-evidences $\ln
\mathcal{Z}_A - \ln \mathcal{Z}_B$, also called the Bayes factor then
quantifies how well A may fit the data when compared with model B.
\citet{Jeffreys} provides a scale on which we can make qualitative conclusions
based on this difference: $\Delta\mbox{ln} \mathcal{Z} < 1$ is not significant,
$1 < \Delta\mbox{ln} \mathcal{Z} < 2.5$ significant, $2.5 < \Delta\mbox{ln}
\mathcal{Z} < 5$ strong and $\Delta\mbox{ln} \mathcal{Z} > 5$ decisive. 
 
\section{Primordial Power Spectrum Parameterisation}
\subsection{H-Z, power-law and running spectra}
\label{h_z}
The early Universe as observed in the CMB is highly homogeneous on large scales
suggesting that any primordial spectrum of density fluctuations should be close
to scale invariant. The H-Z spectrum is described by an amplitude $A$ for which
we assume a uniform prior of $[15,55]\times 10^{-8}$. Slow-roll inflation, given
an exponential potential, predicts a slightly `tilted' power-law
spectrum parameterised as:
\begin{equation}
P(k)=A \left(\frac{k}{k_0}\right)^{n-1},
\label{single_index}
\end{equation}
where the spectral index should be close to unity; we assume a uniform prior on
$n$ of $[0.5,1.5]$; $k_0$ is the pivot scale (set to 0.05 Mpc$^{-1}$) of which
$n$ and the amplitude $A$  are functions. For a generic inflationary potential
we should also account for any scale dependence of $n(k)$ called
\emph{running},  so that to first order:
\begin{equation}
P(k) = A
\left(\frac{k}{k_0}\right)^{n-1+(1/2)\ln (k/k_0)(dn/d\ln k)},
\end{equation}
where $dn/d\ln k$ is the running parameter $n_{run}$; for which we assume a
uniform prior of $[-0.15,0.15]$.

\begin{center}
\begin{figure}
    	\subfigure{
          \includegraphics[width=.25\columnwidth]{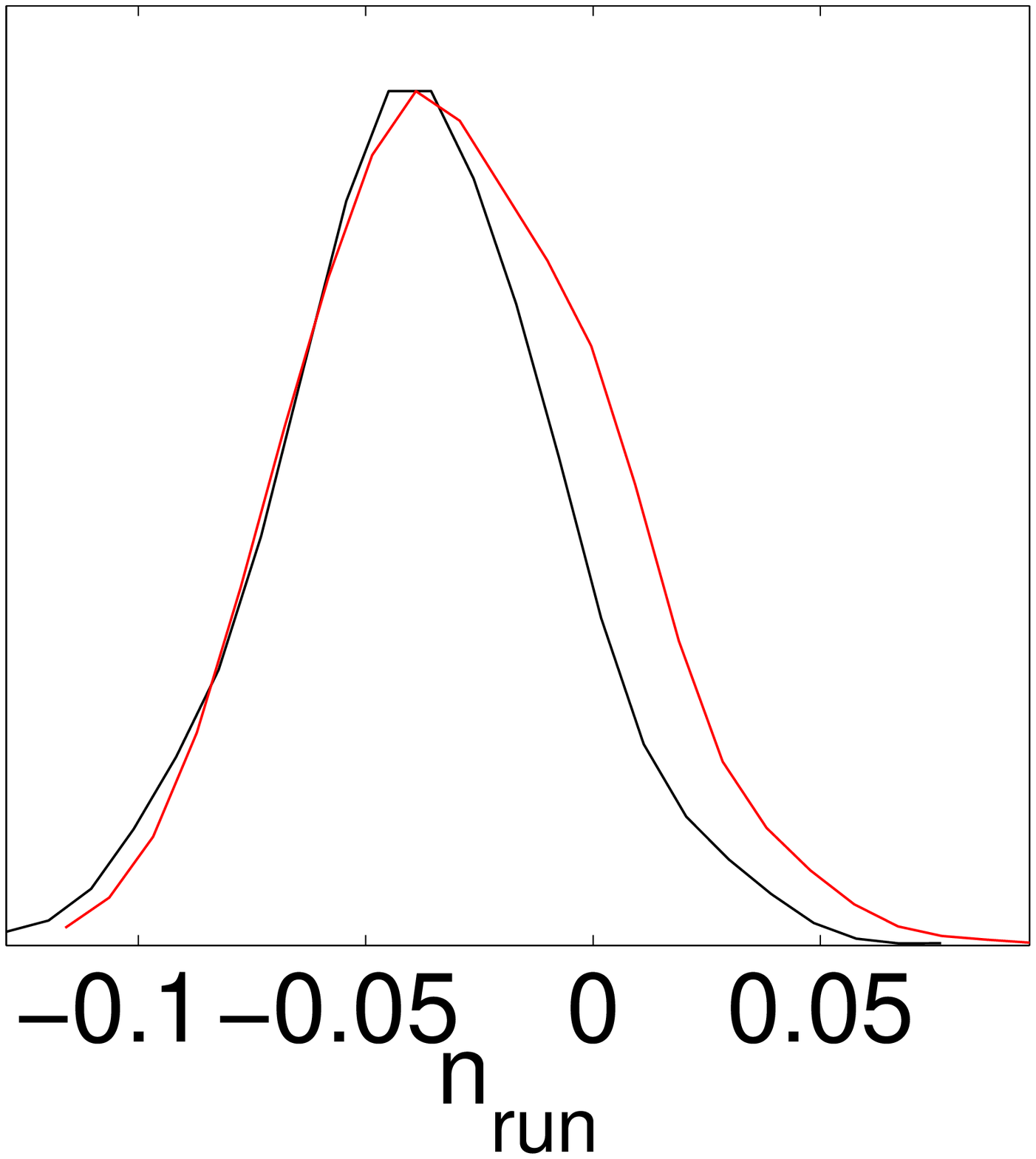}}\\
	\subfigure{
          \includegraphics[width=.25\columnwidth]{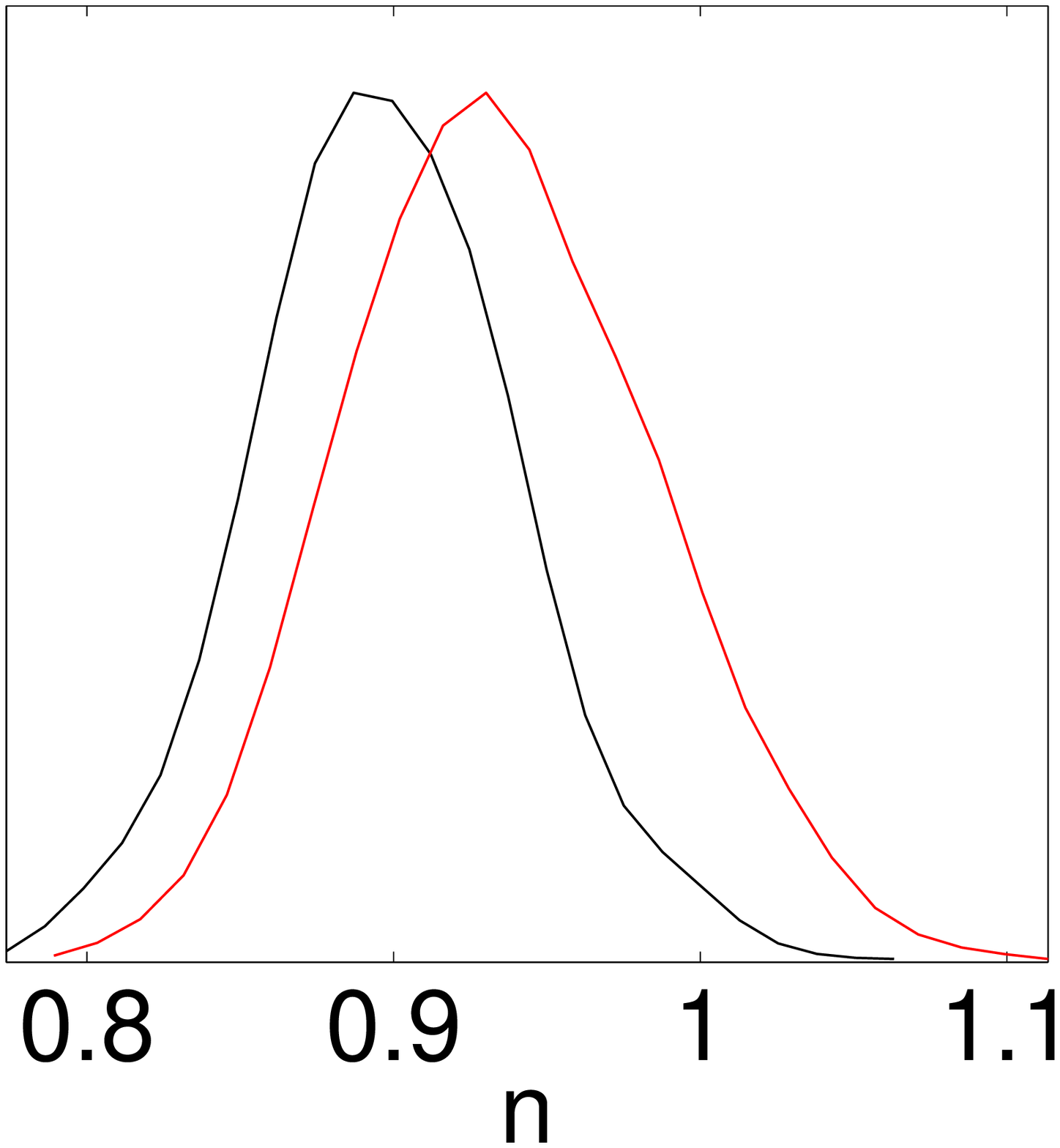}}
	  \hspace{0.2cm}
     	\subfigure{
           \includegraphics[width=.25\columnwidth]{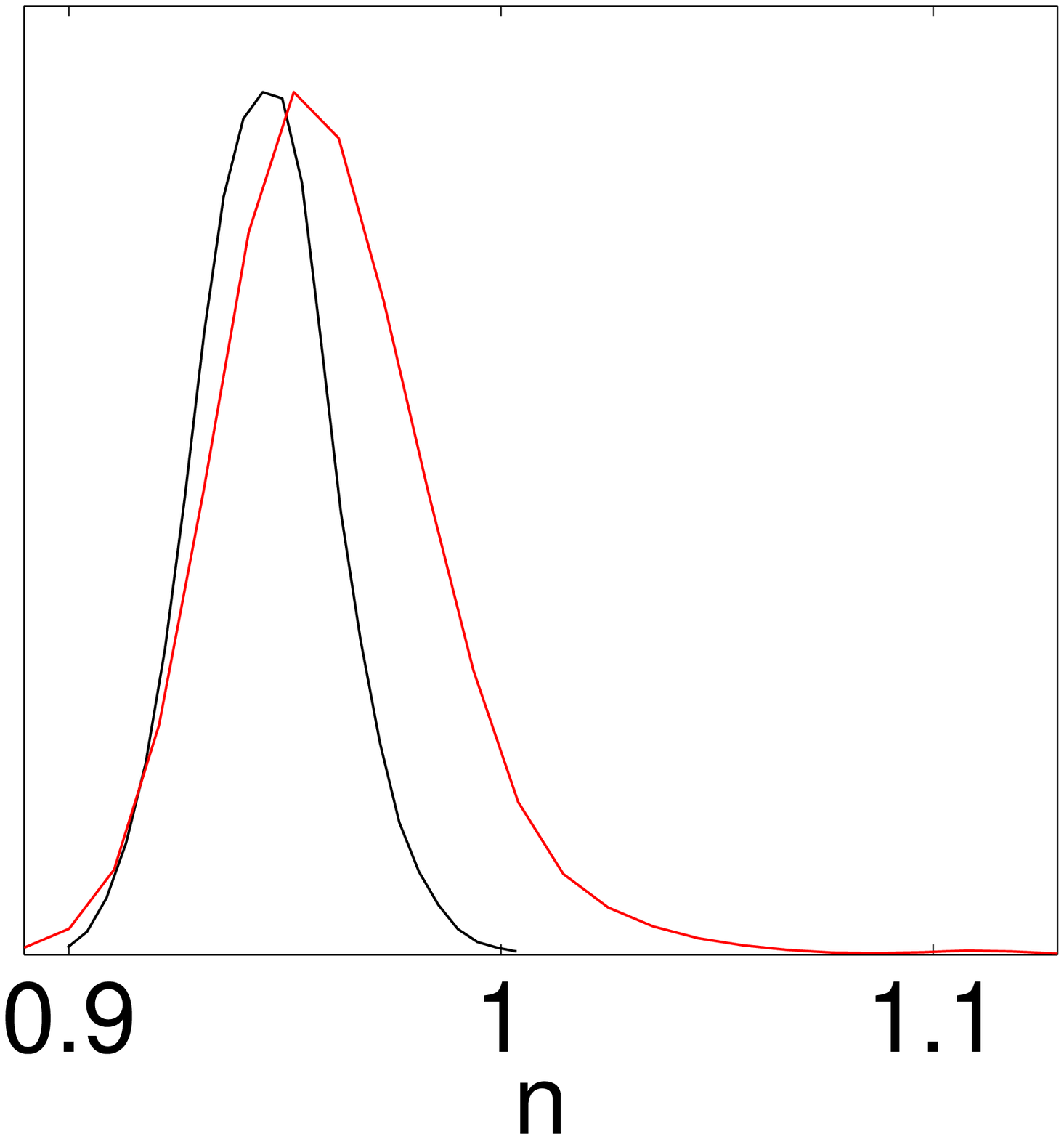}}\\
	\setcounter{subfigure}{0}
	\subfigure[Running]{
           \includegraphics[width=.25\columnwidth]{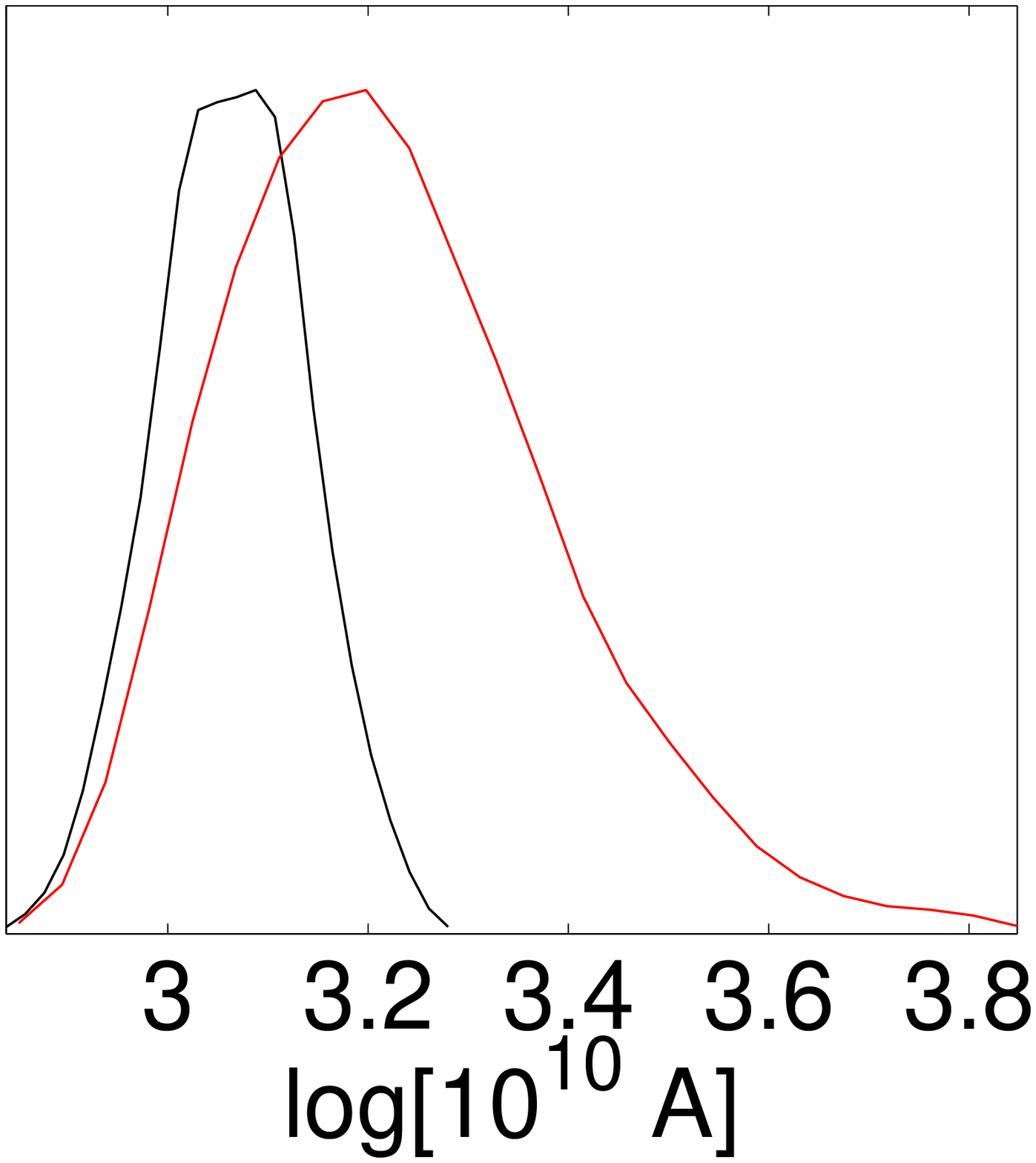}}
	   \hspace{0.2cm}
	\setcounter{subfigure}{1}
	\subfigure[Single-Index]{
           \includegraphics[width=.25\columnwidth]{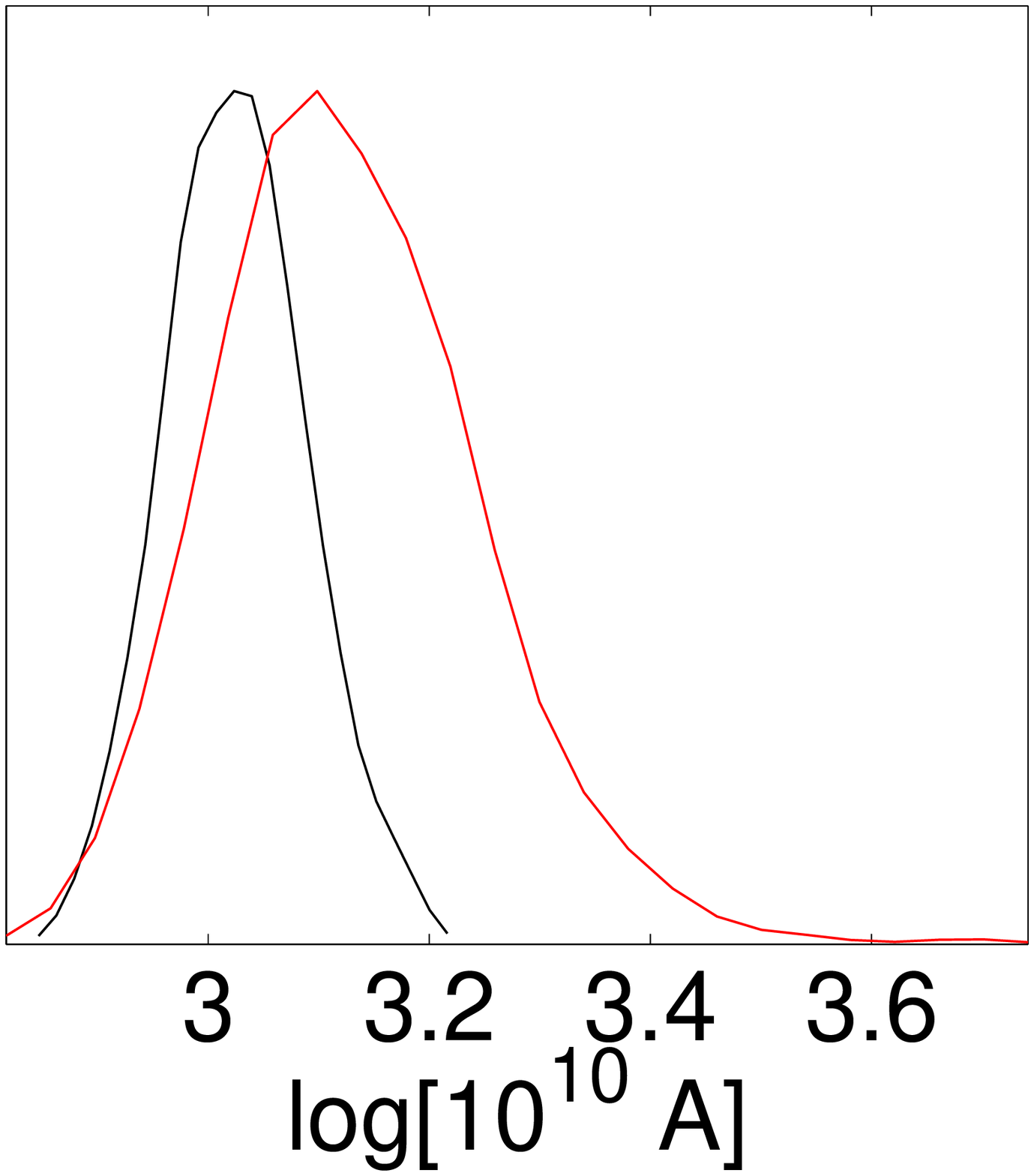}}
	\hspace{0.2cm}
	\setcounter{subfigure}{2}
	\subfigure[H-Z]{
           \includegraphics[width=.25\columnwidth]{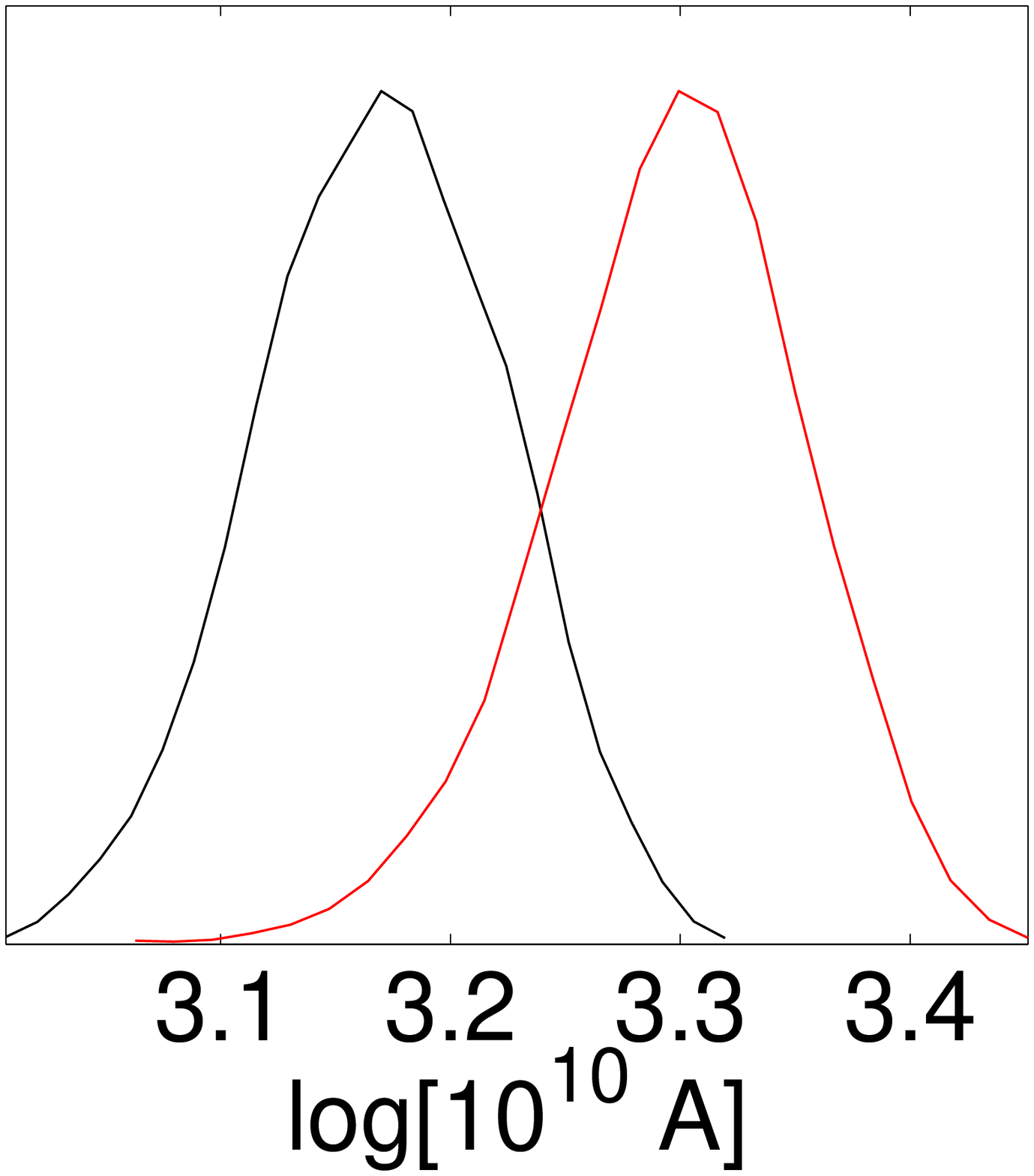}}        	         
\caption{Marginalised parameter constraints with dataset I (black) for the H-Z, single-index and
running models compared with our WMAP1 (Bridges06) analysis (red).}
\label{figure1}
\end{figure}
\end{center}

\begin{center}
\begin{figure}
    	\subfigure{
          \includegraphics[width=.25\columnwidth]{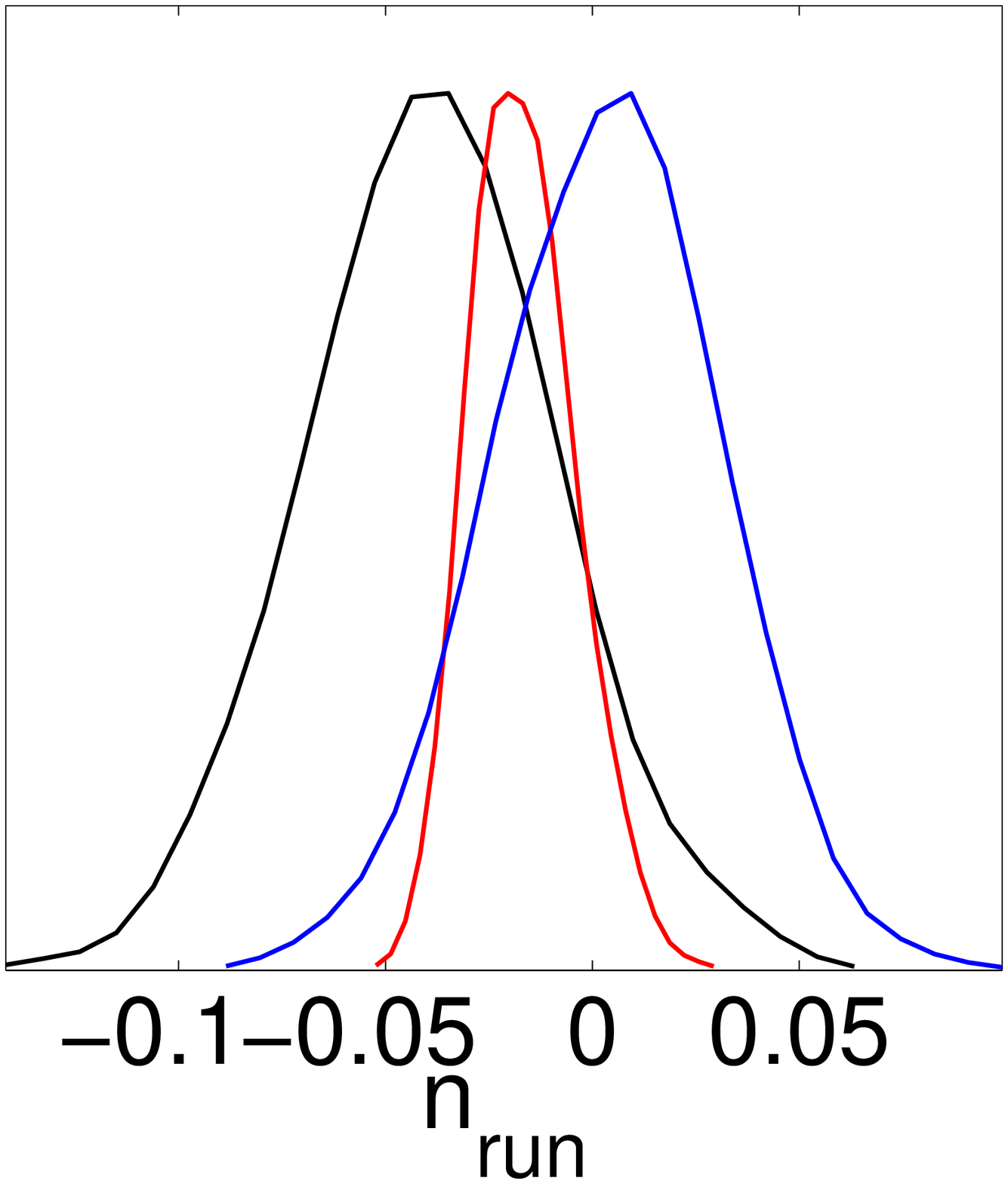}}\\
	\subfigure{
          \includegraphics[width=.25\columnwidth]{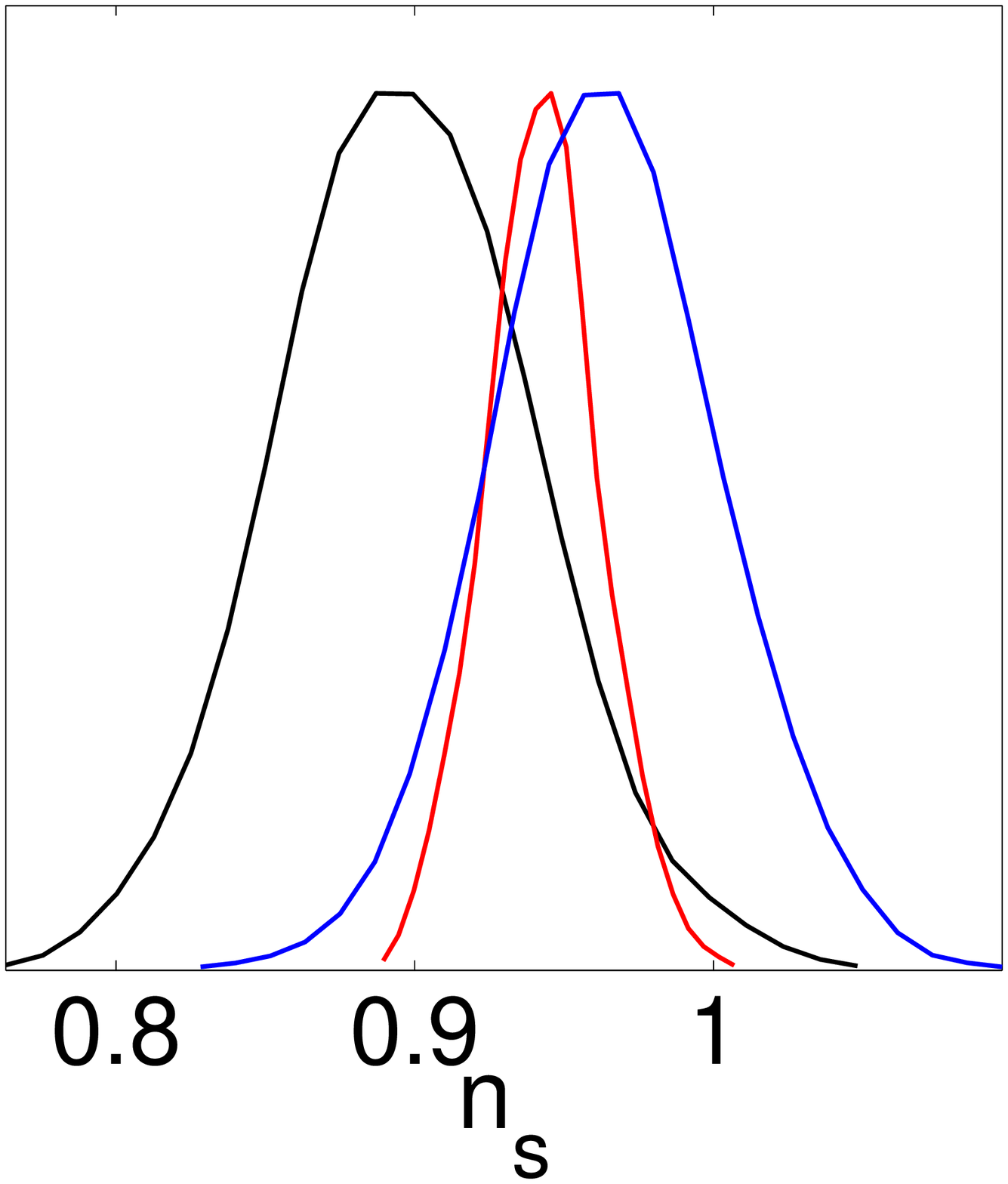}}
	  \hspace{0.2cm}
     	\subfigure{
           \includegraphics[width=.265\columnwidth]{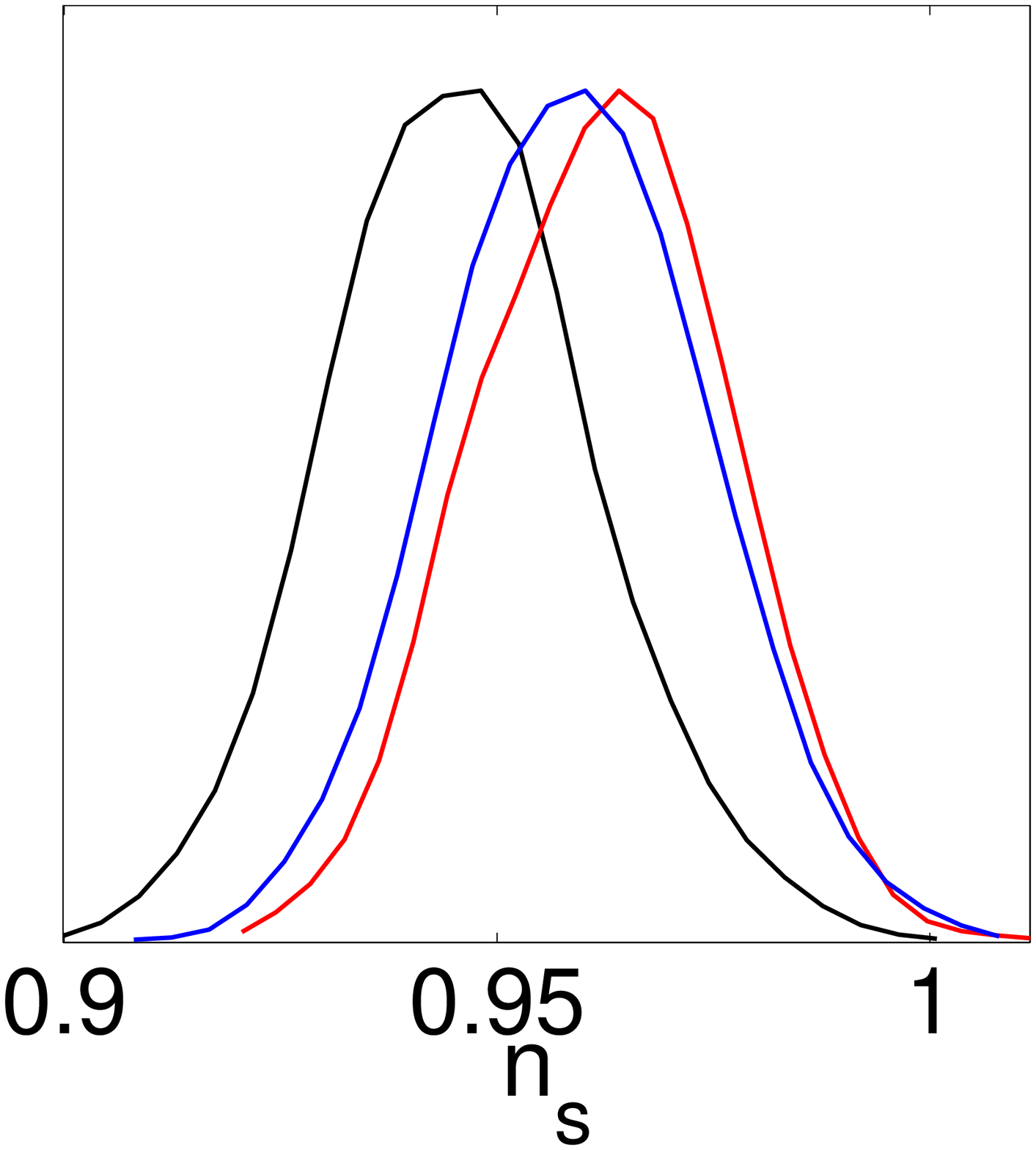}}\\
	\setcounter{subfigure}{0}
	\subfigure[Running]{
           \includegraphics[width=.25\columnwidth]{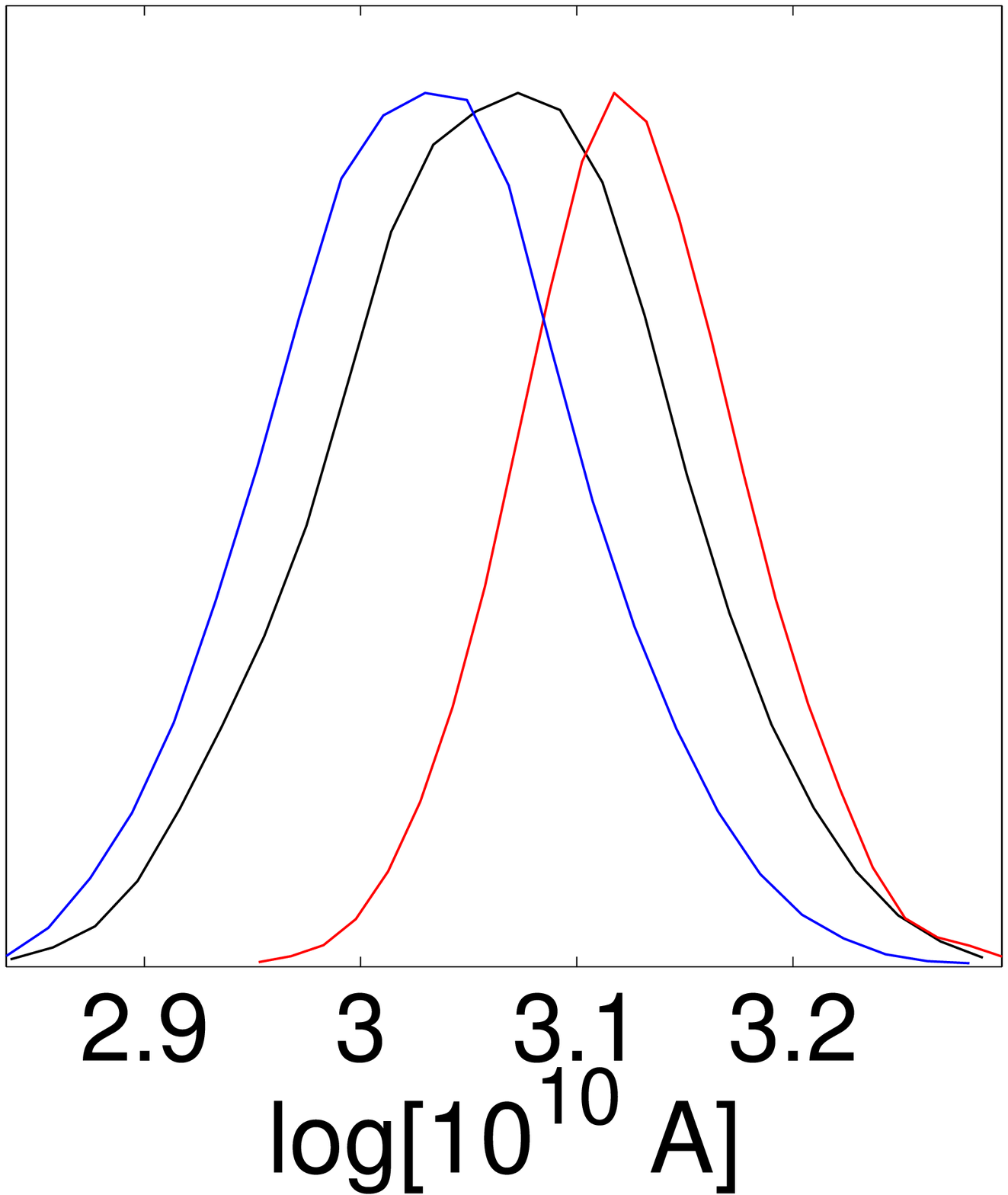}}
	   \hspace{0.2cm}
	\setcounter{subfigure}{1}
	\subfigure[Single-Index]{
           \includegraphics[width=.25\columnwidth]{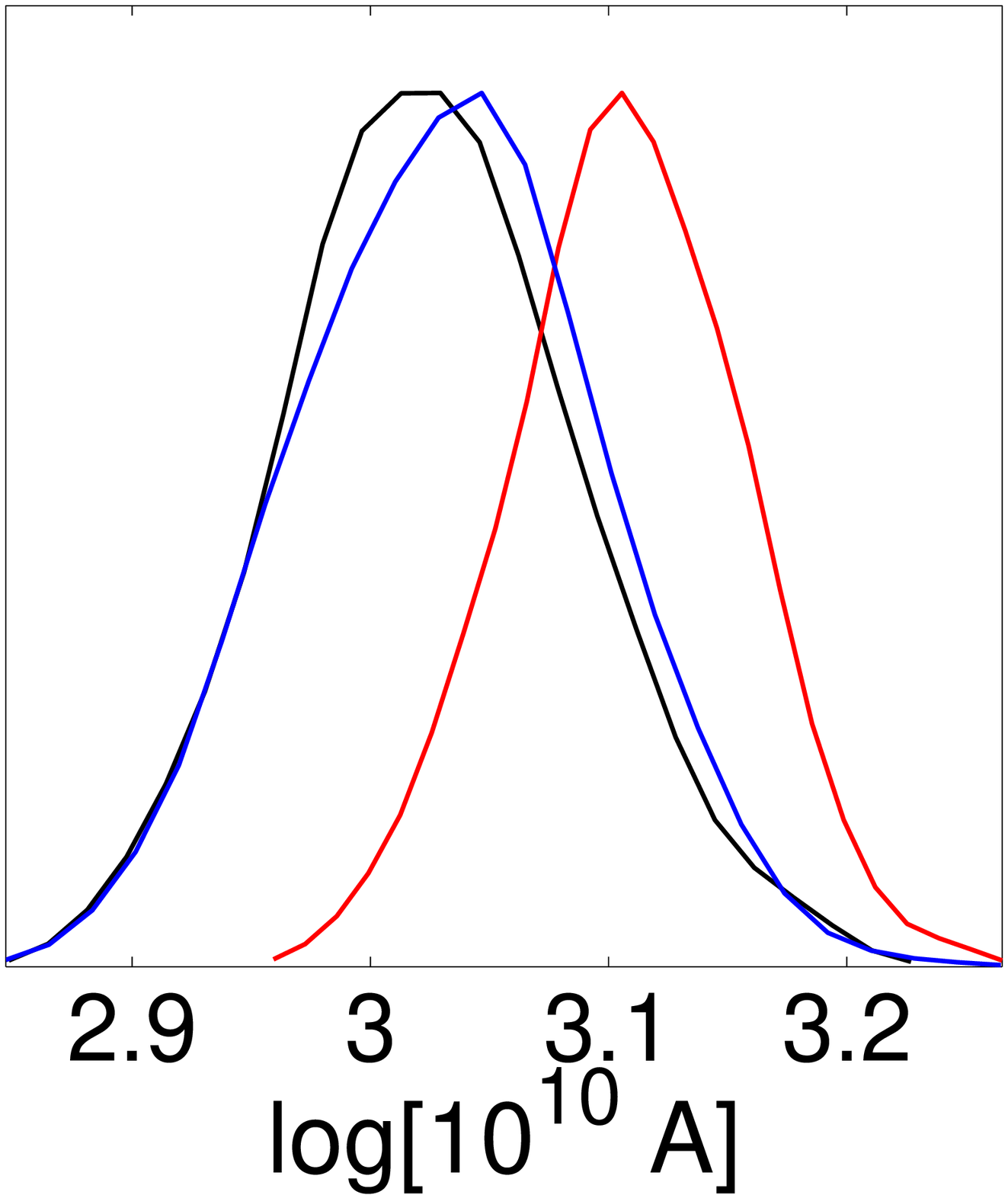}}
	\hspace{0.2cm}
	\setcounter{subfigure}{2}
	\subfigure[H-Z]{
           \includegraphics[width=.25\columnwidth]{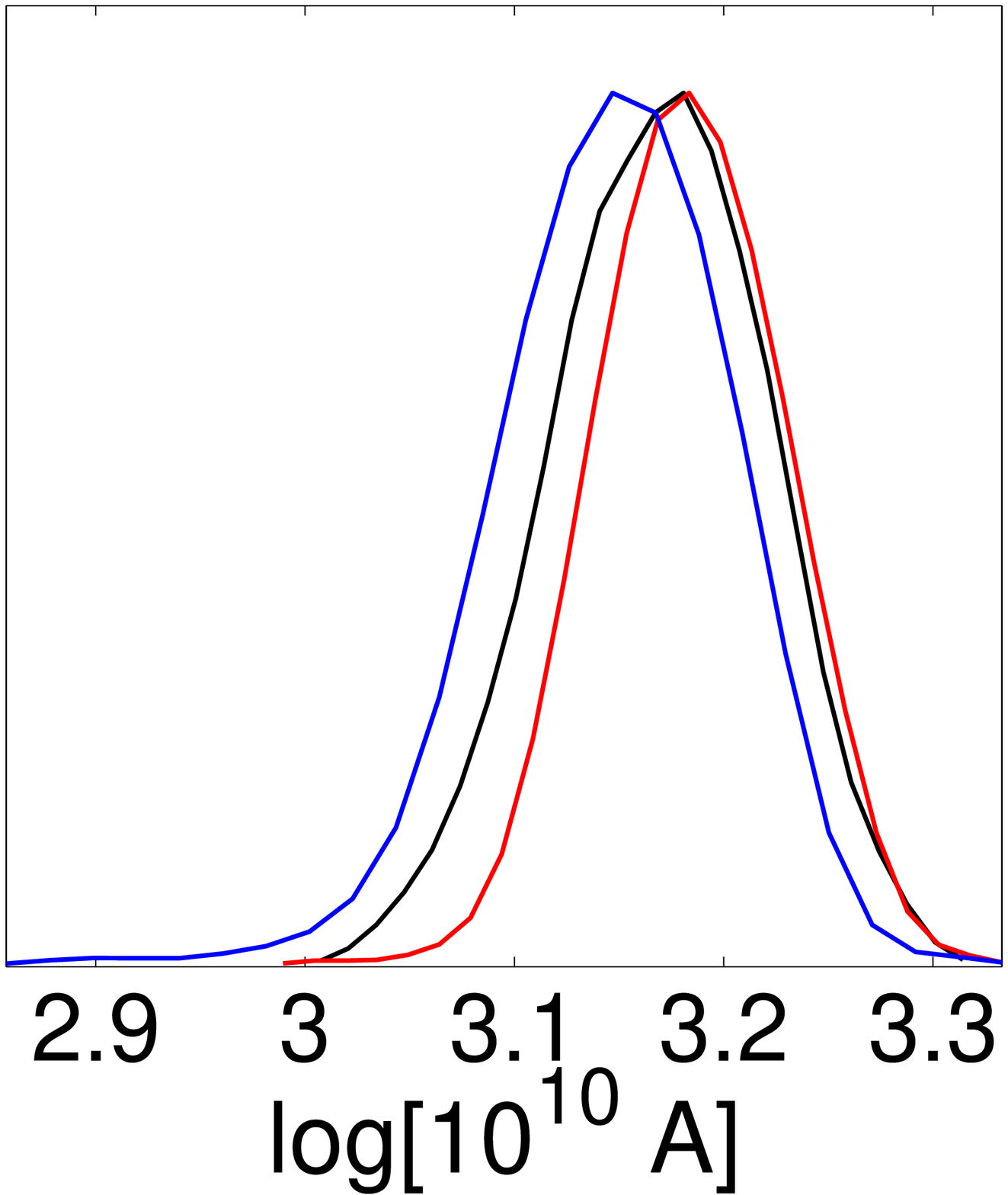}}        	         
\caption{Comparison of parameter constraints using dataset I (black) for the H-Z, single-index and
running models with dataset II (red) and dataset III (blue).}
\label{figure2}
\end{figure}
\end{center}

The inclusion of WMAP3 in dataset I now places impressively tight constraints on all three models (see Fig.
\ref{figure1}). The unexpected reduction in the value of the optical depth to
reionisation, to $\tau \sim 0.09$ has had the effect of reducing the overall
amplitude of the power spectrum, due to the well known $\tau$-$A$ degeneracy. This
effect is noticeable in all cases but most particularly so in the H-Z spectrum in
Fig. \ref{figure1} (c). For the first time the single spectral index model now
exhibits a constraint, to $1\sigma$, of $n_s=0.95 \pm 0.02$ (see Fig.
\ref{figure1} (b)) excluding the possibility of a scale-invariant spectrum
at this confidence level. Furthermore, a pure power-law (with $n_{run} = 0$) is
also excluded at the $1\sigma$ level with $n_{run} = -0.038 \pm 0.030$ (Fig.
\ref{figure1} (a)).  
The addition of Ly-$\alpha$ data further increases constraints on all spectral parameters (see Fig.
\ref{figure2}), particularly so on spectral running. While a scale invariant spectrum  is
also ruled out with this dataset ($n_s = 0.96 \pm 0.02$) a running spectrum is not preferred, with a constraint 
on  $n_{run} = 0.015 \pm 0.015$ representing a doubling in
accuracy over dataset I.  A further tension exists between the amplitude of fluctuations as
found using dataset I alone and when combined with Ly-$\alpha$, the latter preferring a much larger value of $\sigma_8$ and thus
higher scalar fluctuation amplitude. \cite{Seljak} estimates this deviation to be at the 
$2\sigma$ level and treat it as a normal statistical fluctuation and \emph{not} a sign of
some unaccounted systematic flaw in either dataset. Since two independent analyses of WMAP3 + Ly-$\alpha$ and WMAP3 + LRG 
both suggest no significant running, a conservative
conclusion is that the running observed with dataset I is simply a statistical anomaly
albeit at close to 2$\sigma$.

\subsection{Large scale cutoff}
Confirmation by WMAP3 of the large-scale decrement in power at $\ell \sim 2$, close to
the cosmic variance limit, supports the possibility of a spectrum with some form of
cutoff. We reexamine  the case of a sharp cutoff as did the WMAP team, parameterising
the scale at which the power drops to zero, $k_c$ with a choice of prior $[0.0,0.0006]$
Mpc$^{-1}$ which was made to limit the study to regions up to $\ell \sim 6$ at which point
an appreciable cutoff is no longer observed.
\begin{equation}
P(k) = \left\{ \begin{array}{ll}
         0,& \mbox{$k < k_c$}\\
     A \left(\frac{k}{k_0} \right)^{n-1},& \mbox{$k \geq k_c$}\end{array}\right.
\end{equation}
Although this model $P(k)$ is not continuous, as the physical spectrum should be, it
does give an upper limit on the average cutoff scale, which is useful when comparing
for instance the L-D spectrum which does predict a form for the cutoff.   

\begin{center}
\begin{figure}
    	\subfigure{
          \includegraphics[width=.25\columnwidth]{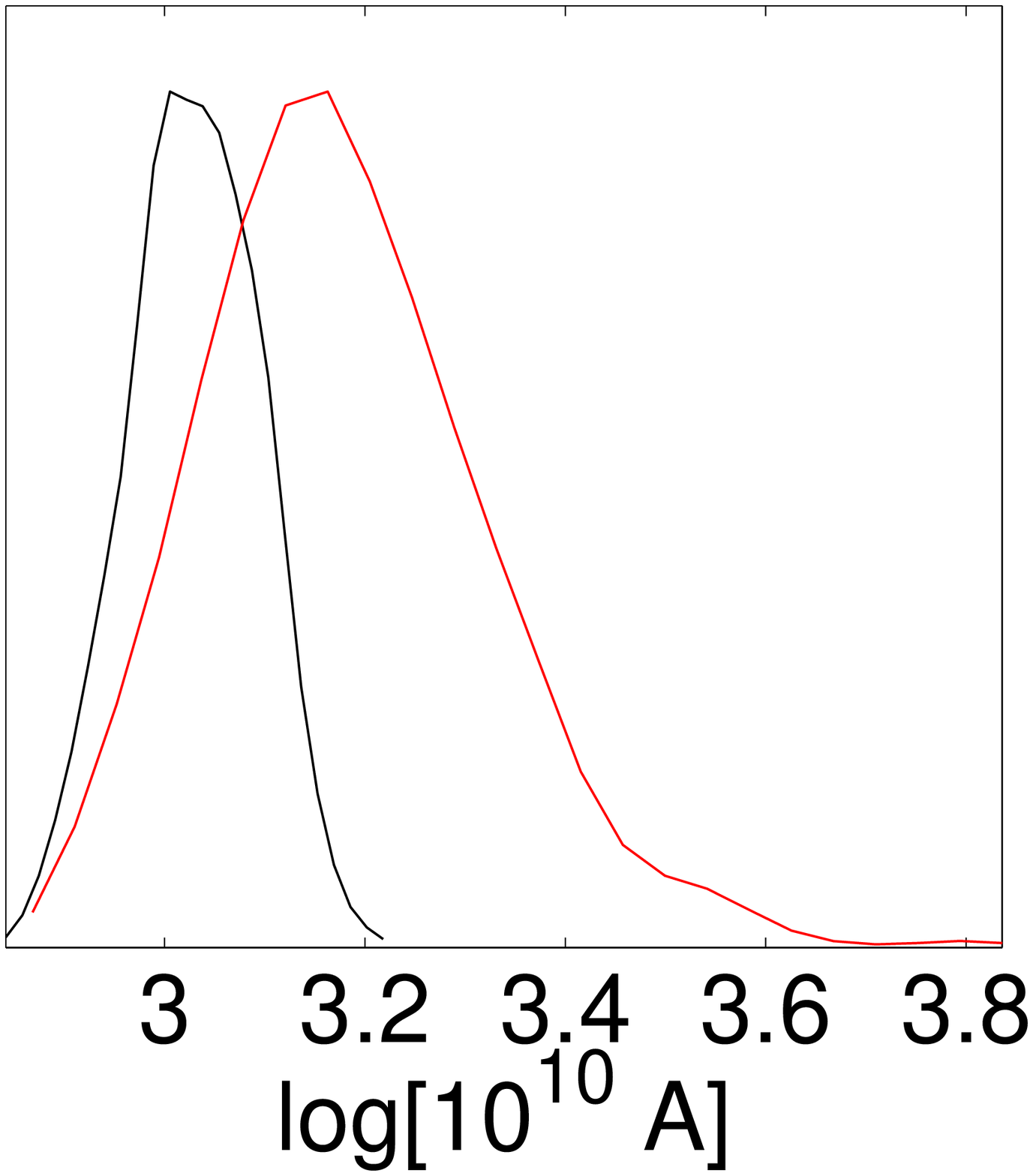}}
     	\hspace{0.2cm}
	\subfigure{
          \includegraphics[width=.25\columnwidth]{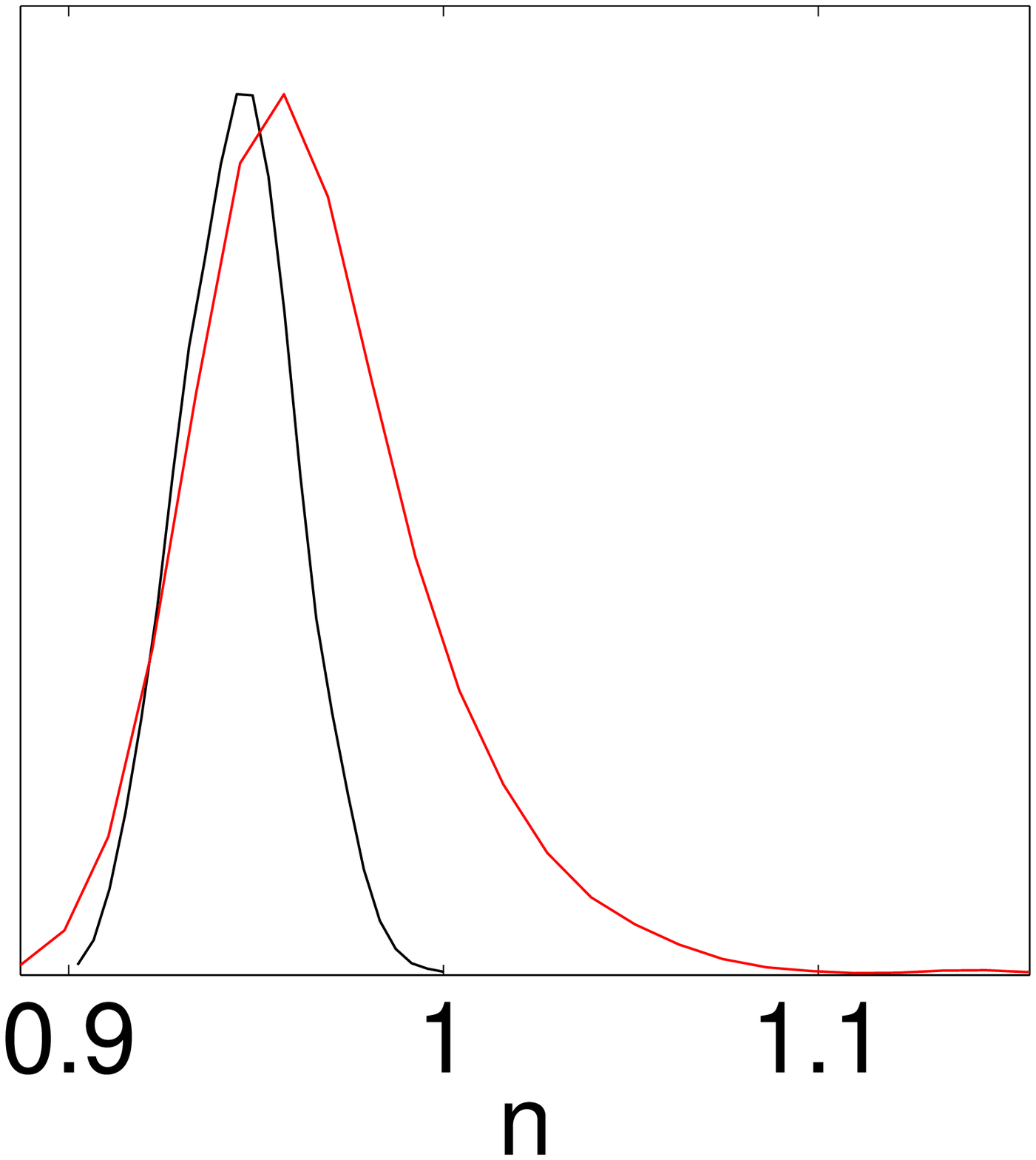}}
	  \hspace{0.3cm}
	 \subfigure{
          \includegraphics[width=.28\columnwidth]{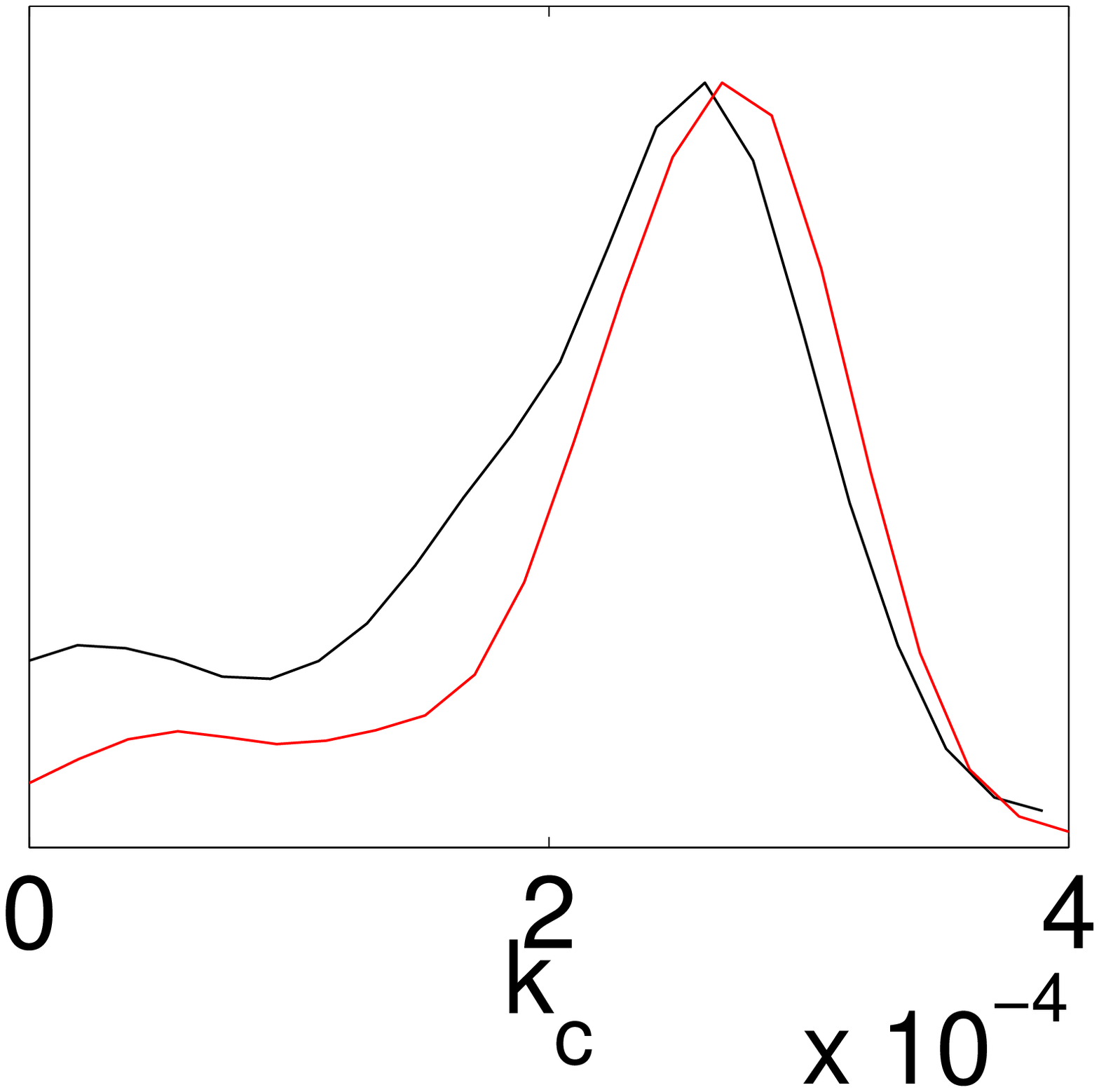}} 	        	         
\caption{Marginalised parameter constraints for dataset I (black) for abrupt cutoff
model compared with our WMAP1 (Bridges06) analysis (red).}
\label{figure3}
\end{figure}
\end{center}

\begin{center}
\begin{figure}
    	\subfigure{
          \includegraphics[width=.25\columnwidth]{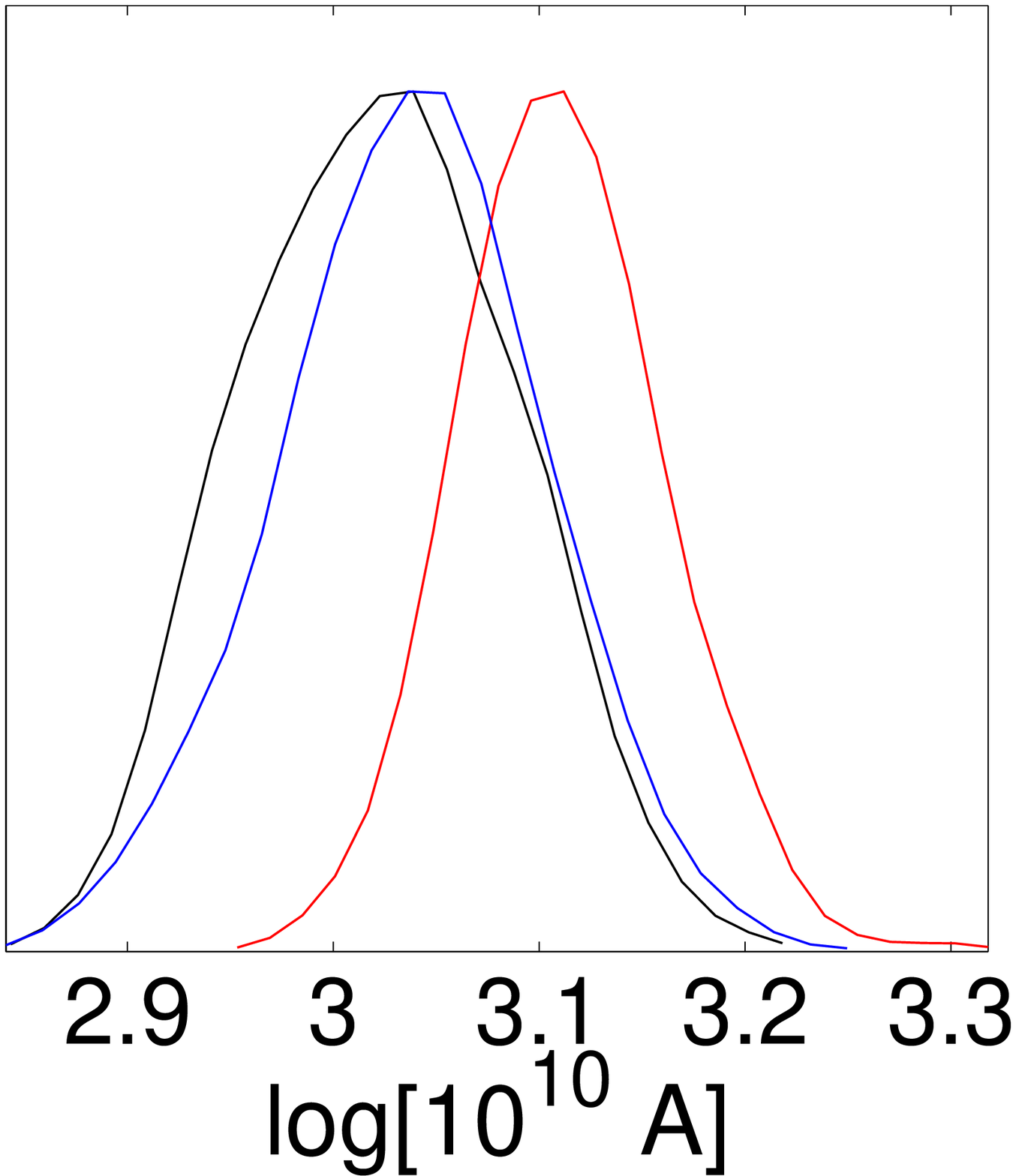}}
     	\hspace{0.2cm}
	\subfigure{
          \includegraphics[width=.25\columnwidth]{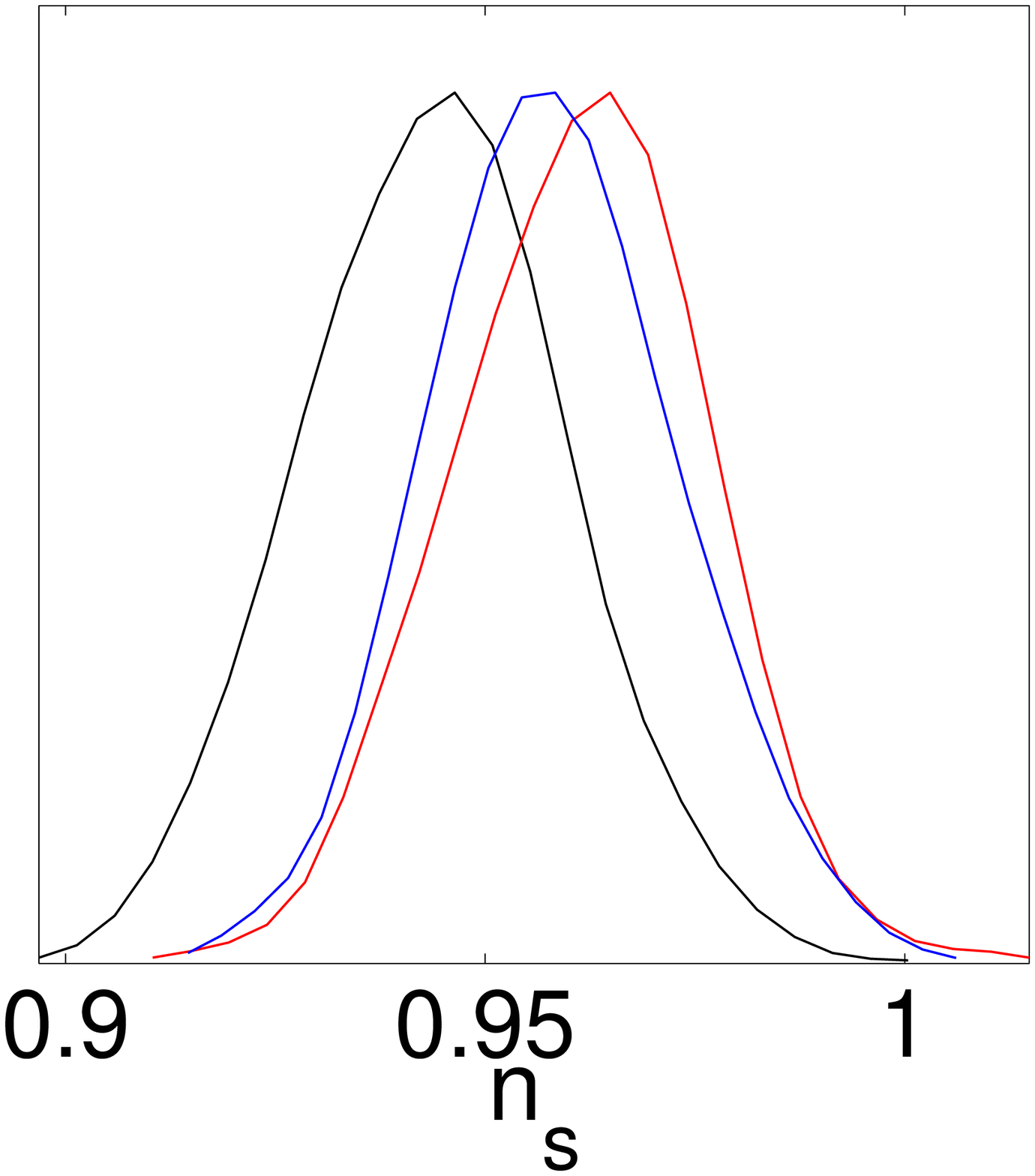}}
	  \hspace{0.3cm}
	 \subfigure{
          \includegraphics[width=.25\columnwidth]{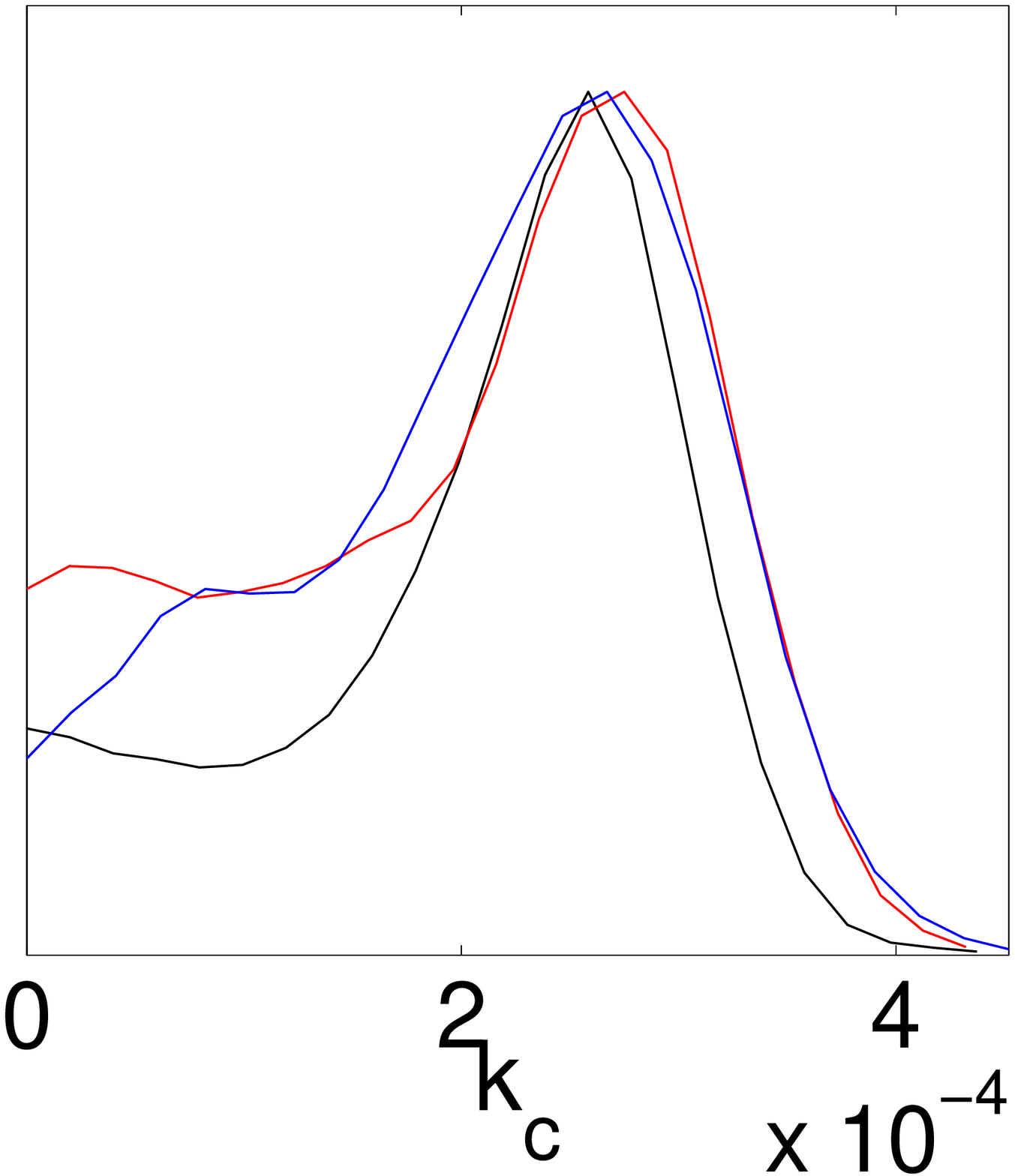}} 	        	         
\caption{Comparison of cutoff constraints using dataset I (black), dataset II (red) and dataset III (blue).}
\label{cutoff_lya_lrg_figure}
\end{figure}
\end{center}

On small scales this spectrum behaves just as the single-index power law and so constraints on
$A$ and $n$ are similar (see Fig. \ref{figure3}). Our WMAP1 constraint on $k_c$ showed a
non-zero likelihood for a cutoff at $k=0$ i.e. no cutoff. This likelihood is marginally lower
with WMAP3 presumably due to the higher amplitude of the $\ell = 3$ multipole which was more
heavily suppressed in WMAP1  observations. The peak in the likelihood encouragingly remains at
around $k_c=2.8\times 10^{-4}$ Mpc$^{-1}$. The scale of this cutoff is much larger than
anything probed by either Ly-$\alpha$ (in dataset II) or LRG (in dataset III), so neither
dataset improves on this scale constraint (see Fig. \ref{cutoff_lya_lrg_figure}). 

\subsection{Broken Spectrum}
Multiple field inflation would produce a symmetry breaking phase transition in
the early universe causing the mass of the inflaton field to change suddenly,
momentarily violating the slow-roll conditions \citep{Adams}. The resultant
primordial spectrum would be roughly scale invariant initially, followed by a
sudden break lasting roughly 1 $e$-fold before returning to scale invariance.
Because the slow-roll conditions are violated it is not trivial to calculate the
form of the break, however a robust expectation is that it will be sharp as the
field undergoing the phase transition evolves exponentially fast to its minimum
\citep{barriga}. We parameterise the spectrum as:
\begin{equation}
P(k) = \left\{ \begin{array}{ll}
         A,& \mbox{$k \leq k_s$}\\
     Ck^{\alpha-1}, & \mbox{$k_s < k \leq k_e$}\\
	B, & \mbox{$k > k_e$} \end{array}\right.,
\end{equation}
where the values of $C$ and $\alpha$ are chosen to ensure continuity. Four power spectrum
parameters were varied in this model: the ratio of amplitudes before and after the break $A/B$ with
prior $[0.3,7.2]$; $k_s$ indicating the start of the break with prior $[0.01,0.1]$ Mpc$^{-1}$;
$\ln(k_e/k_s)$ to constrain the length of the break with prior $[0,4]$ and normalisation $A$ with
prior $[14.9,54.6]\times 10^{-8}$. These represent a very conservative set of priors that allow the model a large degree
of freedom in both the position of the break (which could occur anywhere from $k \sim 0.01$, well
above any possible large scale cutoff) and the form, which could be extended,
so as to mimic a tilted spectrum or occur as a sharp drop. In addition 
a prior that $k_e$ could not exceed 0.1 Mpc$^{-1}$ was imposed so that only a region well covered by
the datasets used was explored. 
 
\begin{center}
\begin{figure}
	\hspace{0.5cm}
    	\subfigure{
          \includegraphics[width=.25\columnwidth]{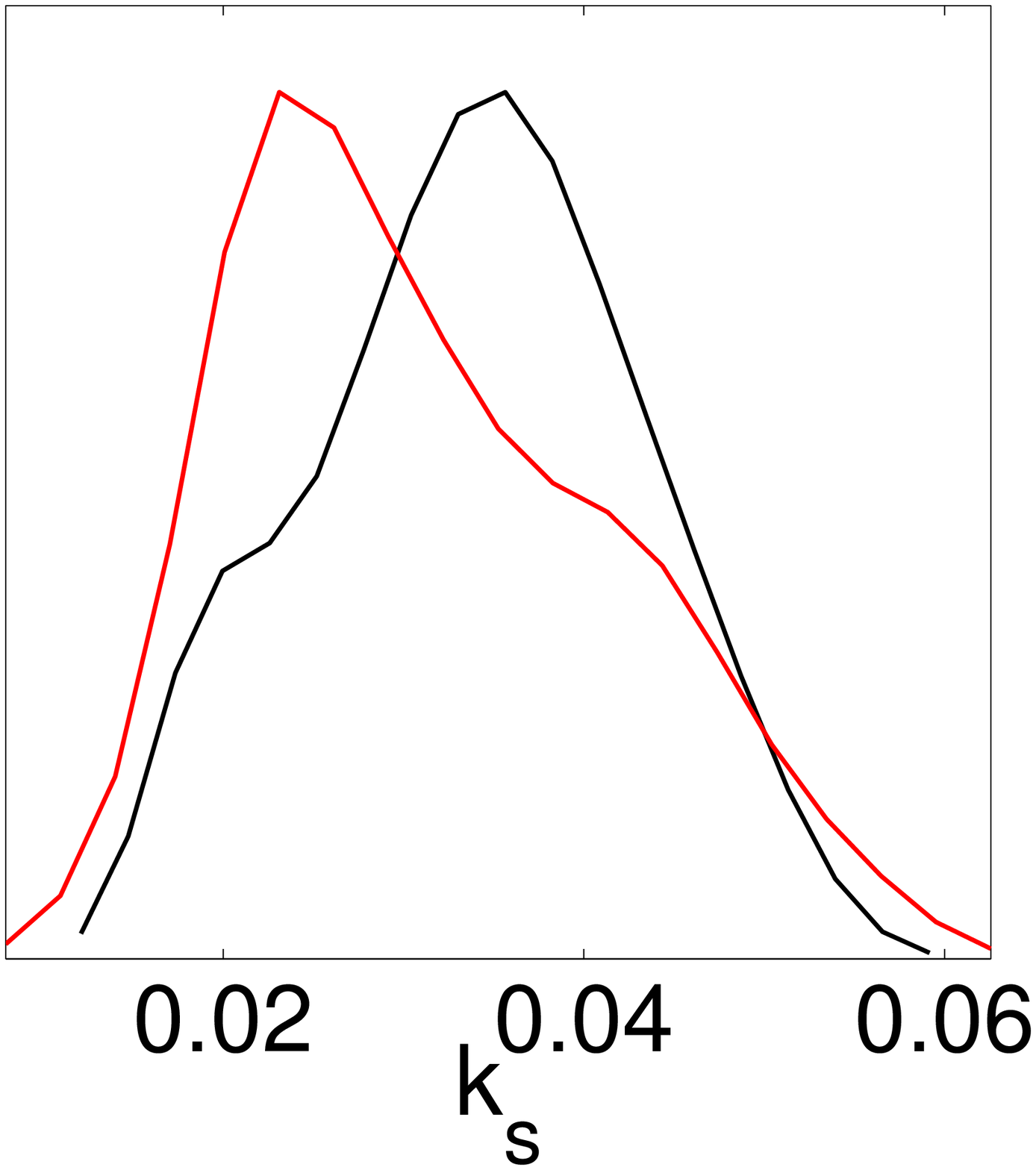}}\\
	\subfigure{
          \includegraphics[width=.35\columnwidth]{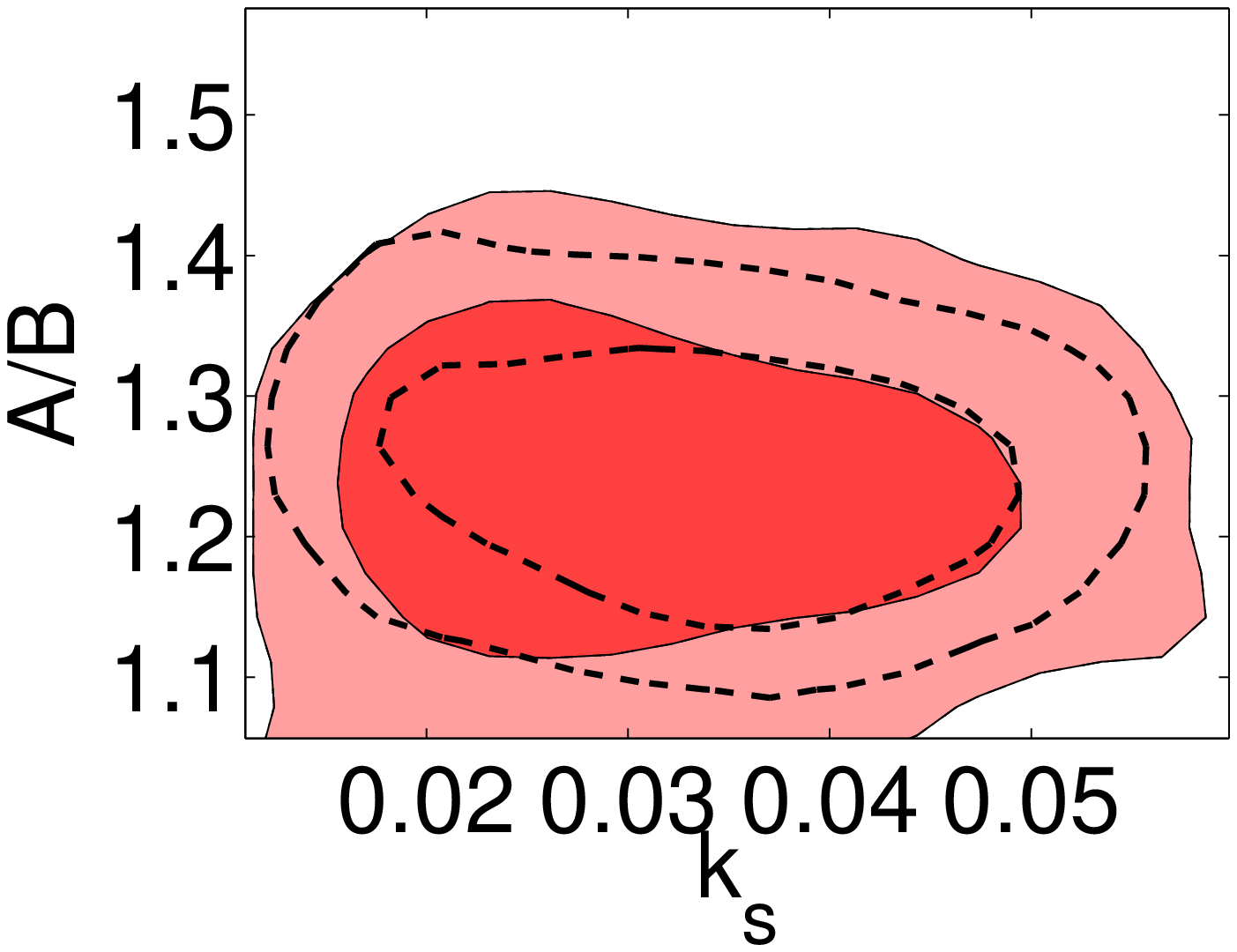}}
     	\hspace{0.5cm}
	\subfigure{
          \includegraphics[width=.25\columnwidth]{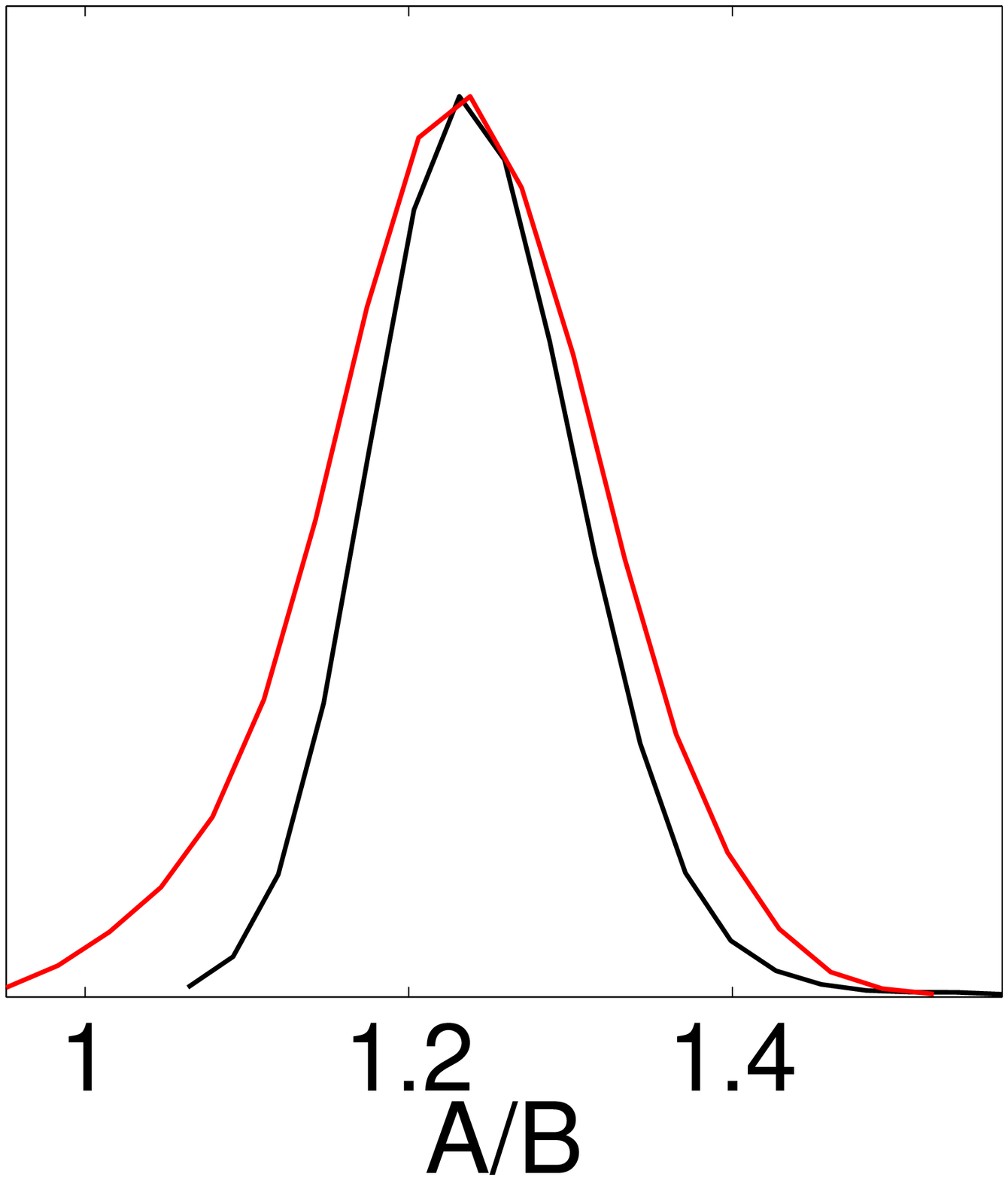}}\\
     	\subfigure{
           \includegraphics[width=.35\columnwidth]{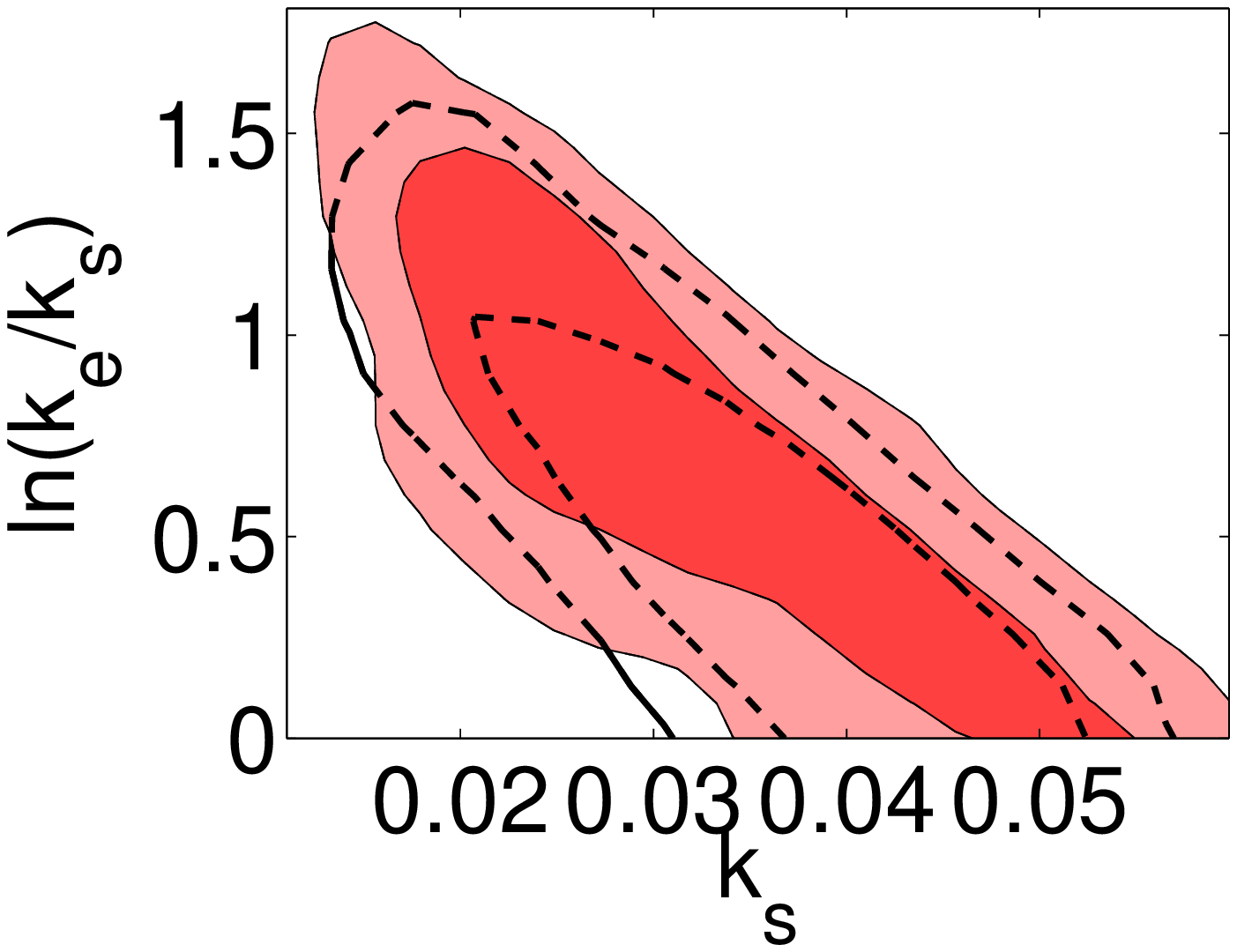}}
	\subfigure{
           \includegraphics[width=.35\columnwidth]{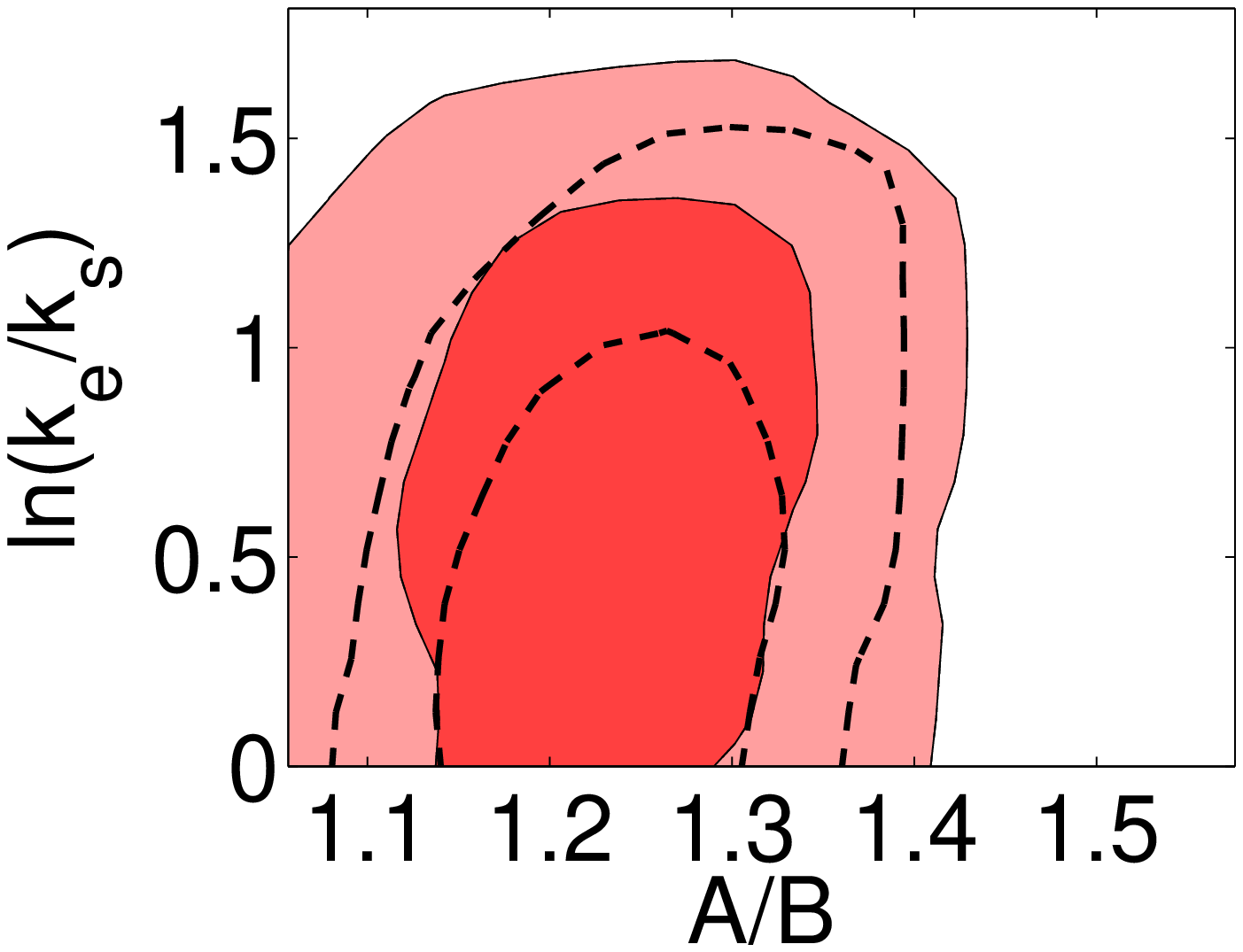}}
	\subfigure{
           \includegraphics[width=.24\columnwidth]{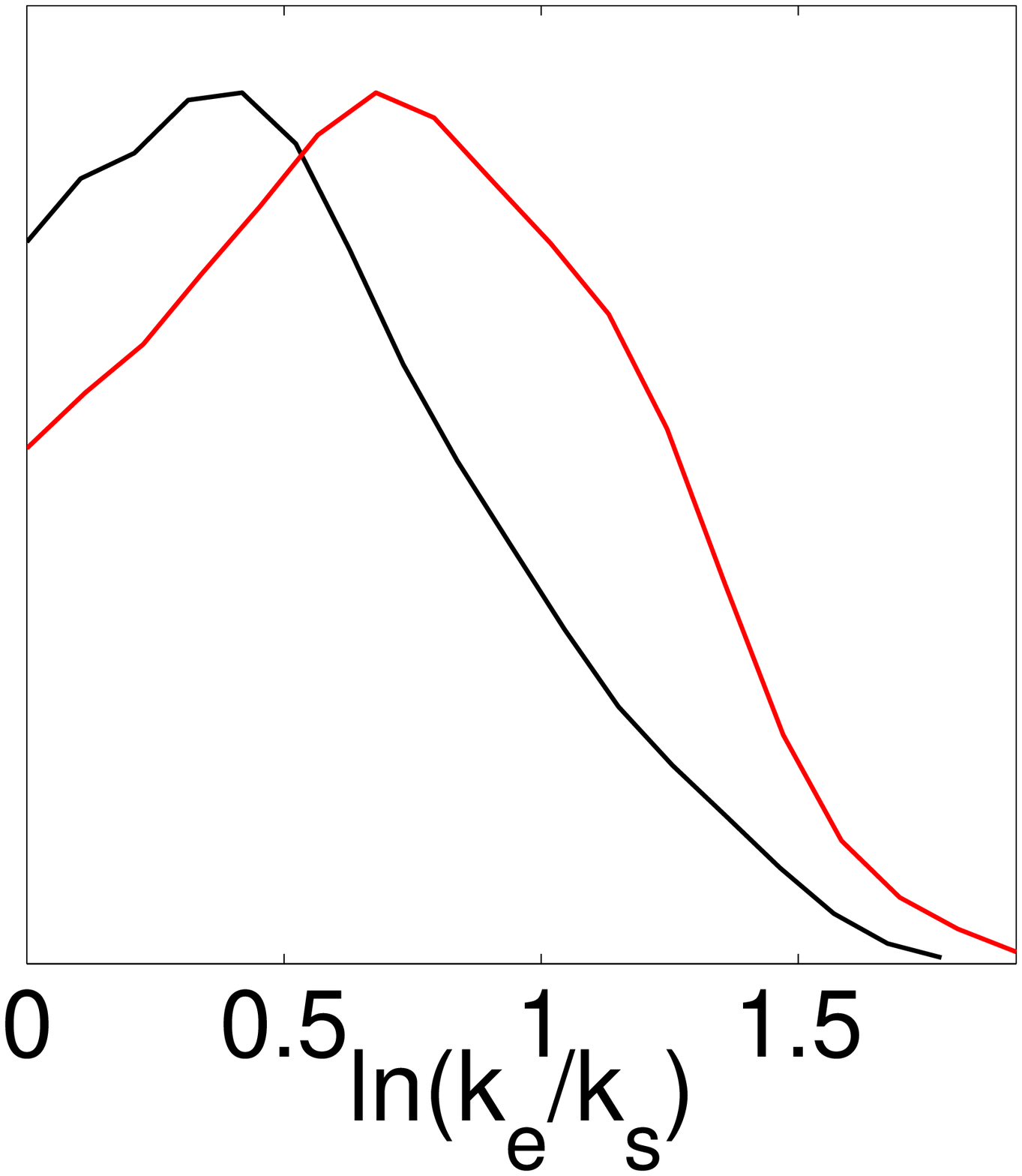}}        	         
\caption{Marginalised 1D and 2D probability constraints for the broken spectrum model, for $k_s$, $\ln(k_e/k_s)$ and $A/B$
for our WMAP1 (Bridges06) analysis (red) and dataset I (black).
2D constraints plotted with $1\sigma$ and $2\sigma$ confidence contours.}
\label{figure4}
\end{figure}
\end{center}

\begin{center}
\begin{figure}
	\hspace{0.5cm}
    	\subfigure{
          \includegraphics[width=.25\columnwidth]{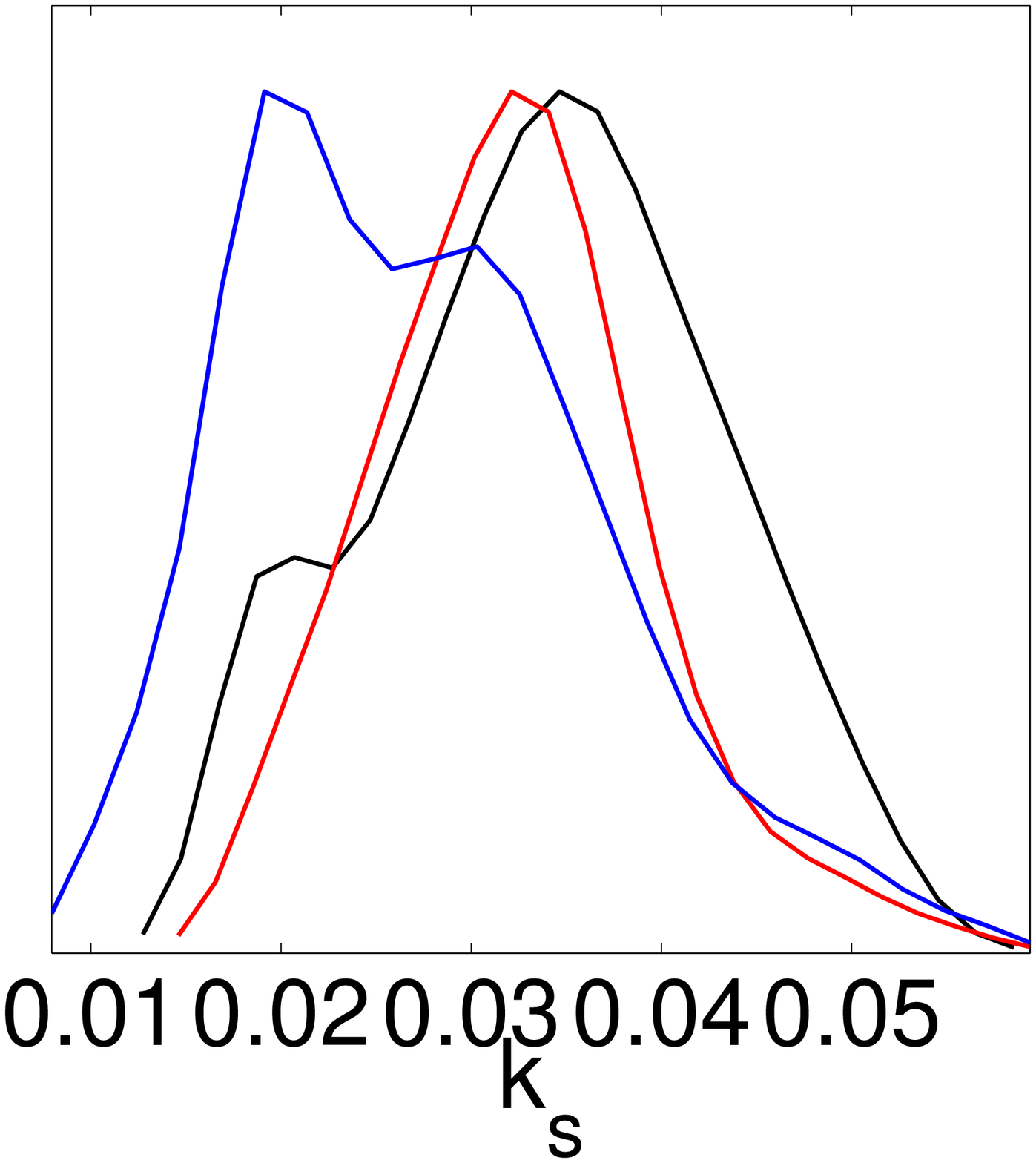}}\\
	\subfigure{
          \includegraphics[width=.35\columnwidth]{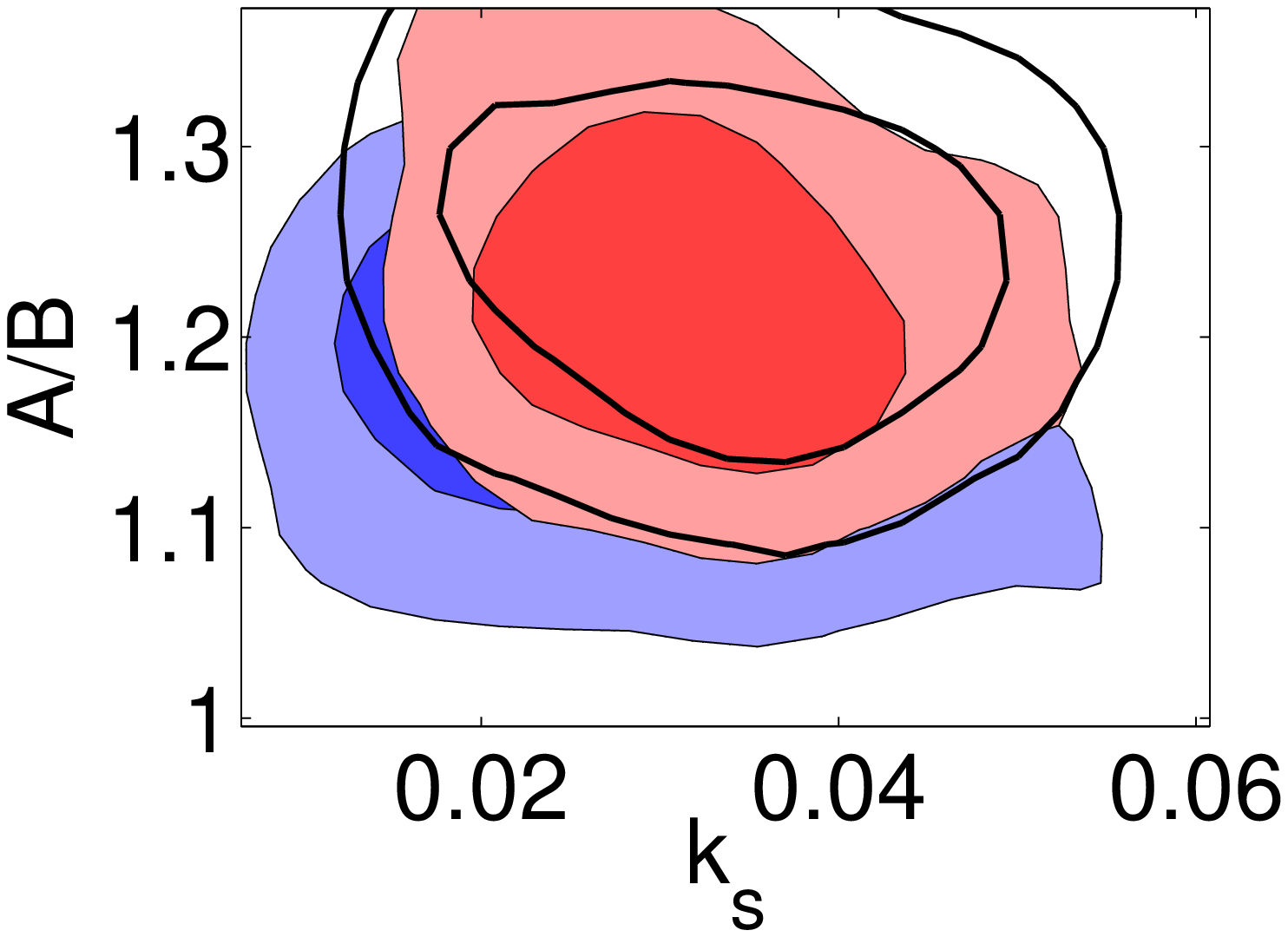}}
     	\hspace{0.5cm}
	\subfigure{
          \includegraphics[width=.25\columnwidth]{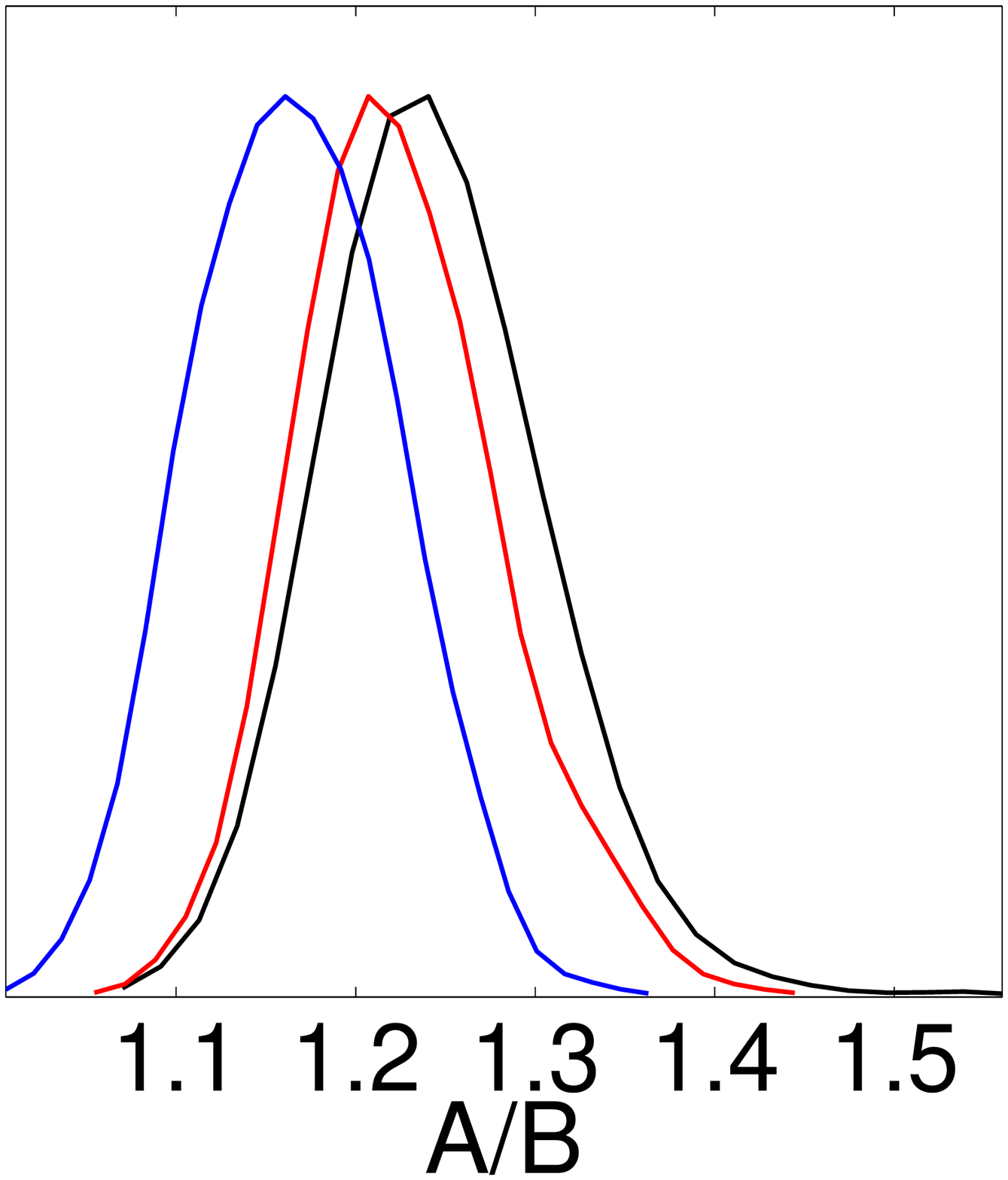}}\\
     	\subfigure{
           \includegraphics[width=.35\columnwidth]{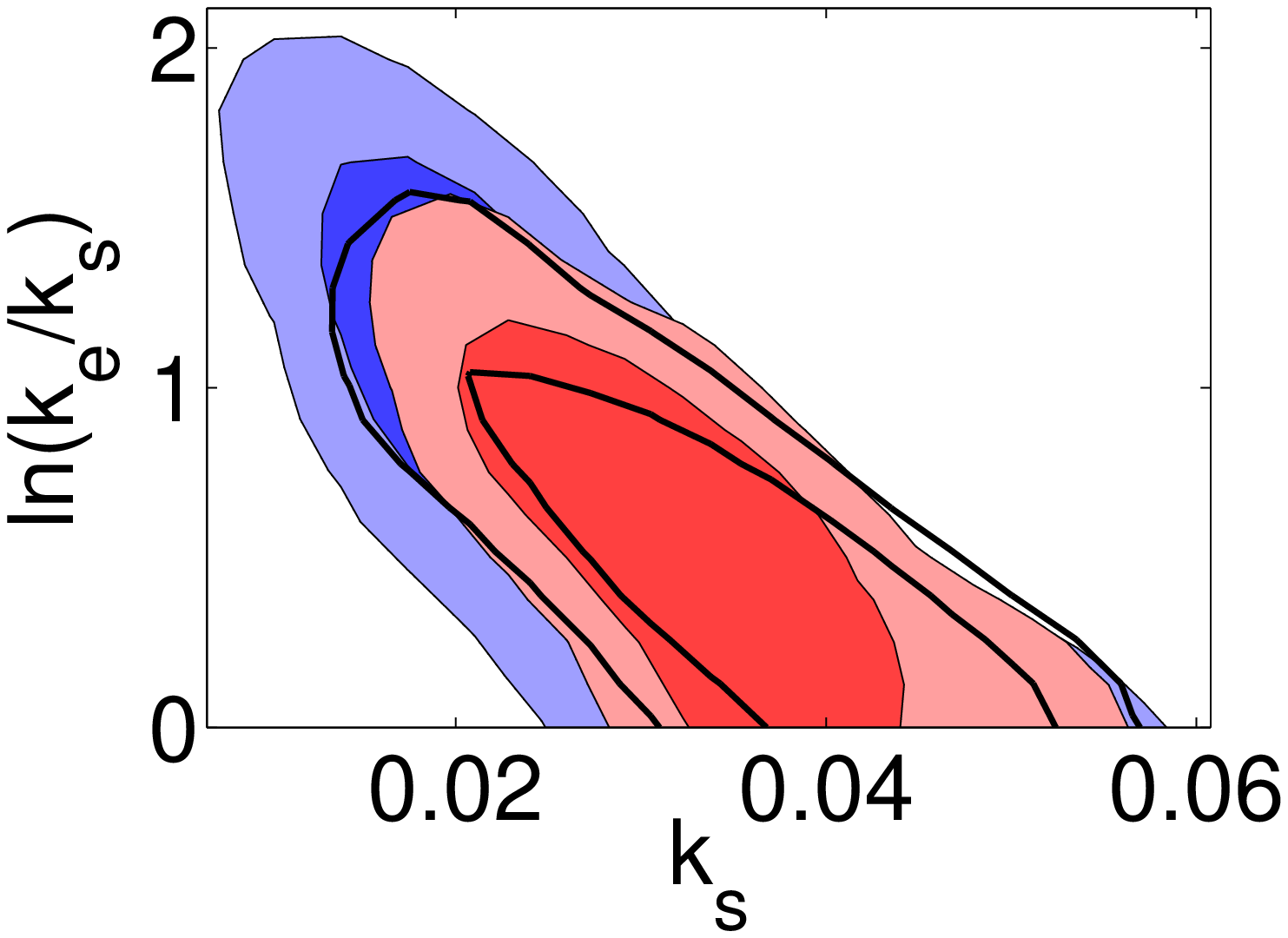}}
	\subfigure{
           \includegraphics[width=.35\columnwidth]{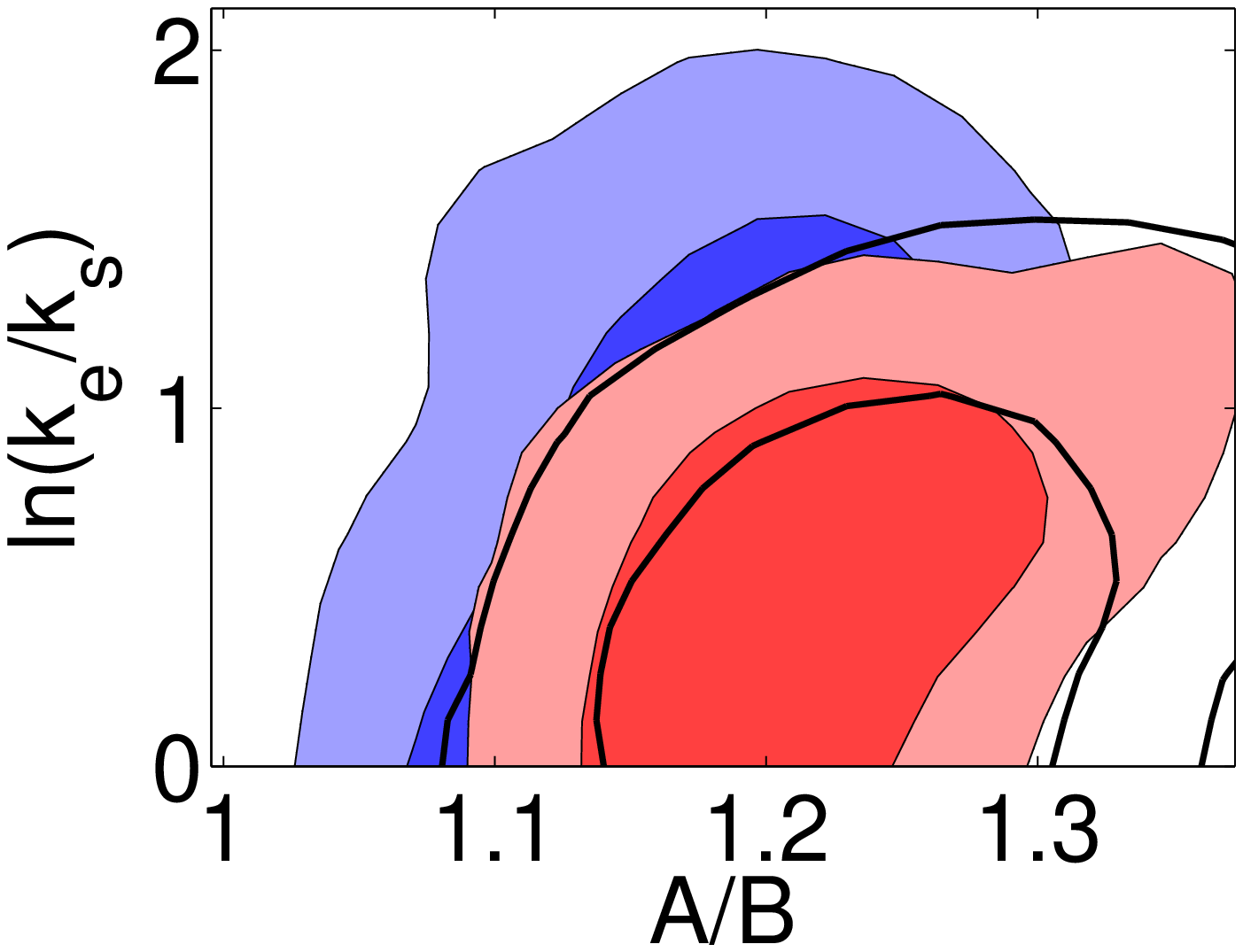}}
	\subfigure{
           \includegraphics[width=.24\columnwidth]{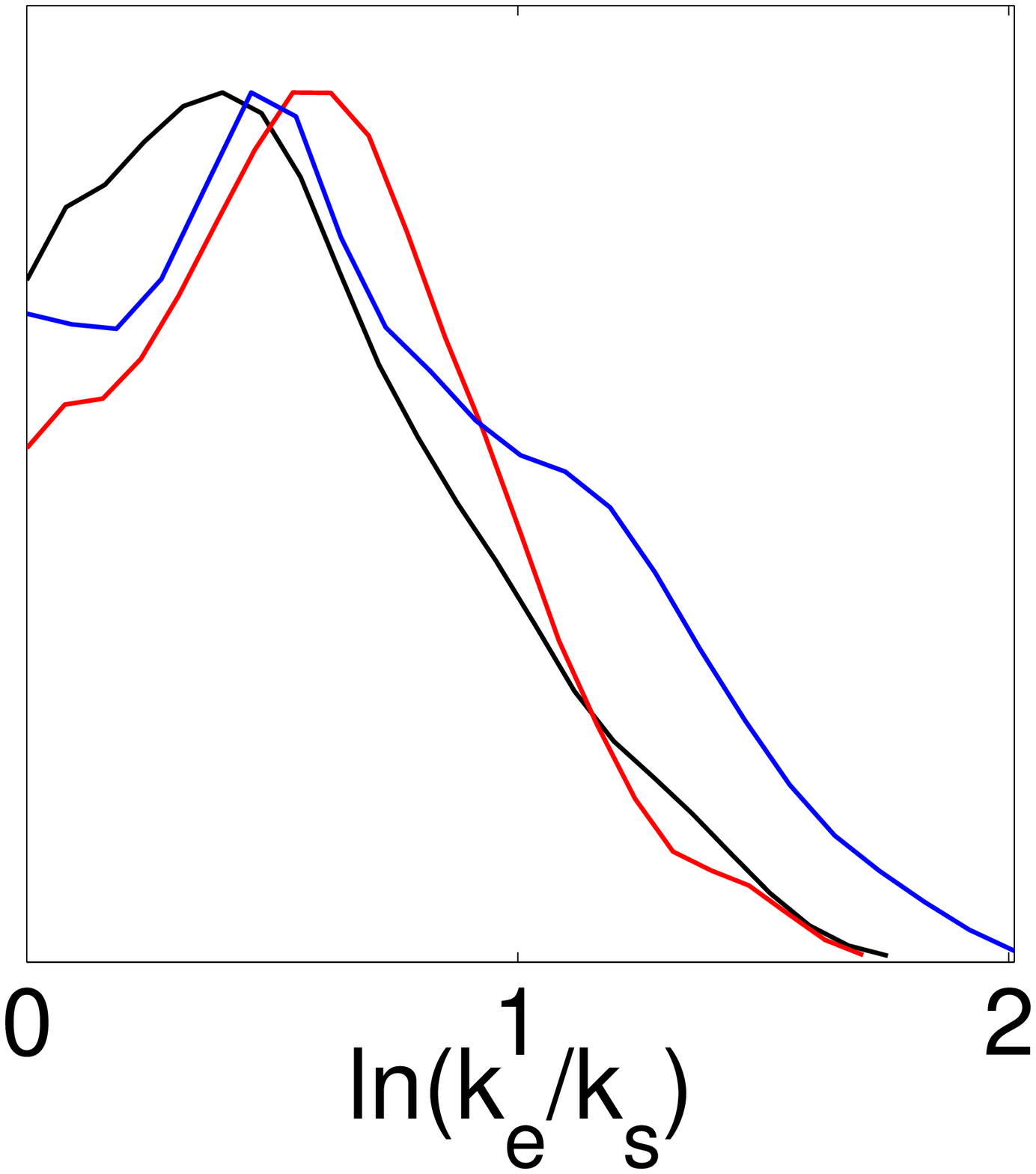}}        	         
\caption{Marginalised 1D and 2D probability constraints for the broken spectrum model, for $k_s$, $\ln(k_e/k_s)$ and $A/B$
for dataset I (black), dataset II (red) and dataset III (blue).
2D constraints plotted with $1\sigma$ and $2\sigma$ confidence contours.}
\label{figure5}
\end{figure}
\end{center}

In this parameterisation a scale invariant spectrum would have an amplitude ratio $A/B = 1$
and a large value of $\ln(k_e/k_s)$. Neither our WMAP1 analysis nor dataset I suggest this to be a
plausible explanation (see Fig. \ref{figure4}) with a distinct drop in amplitude ($A/B \sim
1.2$) starting on scales below $k_s \sim 0.01$ Mpc$^{-1}$.  The bimodal distribution in $\ln
(k_e/k_s)$ vs. $k_s$ observed with WMAP1 is preserved, though less pronounced, with dataset I.
This implies a preference for both a sharp break at a single scale of $k\approx 0.04$
Mpc$^{-1}$ and an extended break begining at $k\approx 0.01$. A drop in power is clearly a
crude way of approximating a tilted spectrum, so the extended break was an expected result.
The sharp break, could be indicative of a some early universe physics or it could simply be
as \citet{Recon} suggest in their WMAP1 analysis, an artifact of the inclusion of the large
scale structure datasets. The transition scale lies close to the point at which large scale
structure data becomes statistically significant. Below this scale the WMAP data would
disfavour a break and above it large scale structure data would. The inclusion of both Ly-$\alpha$ and
LRG (see Fig. \ref{figure5}) data shifts the start of the break to larger scales (most
pronounced with LRG),  but simultaneously lowers the amplitude of the break, thus producing
a more gradual extended slope, providing a better approximation to a tilted spectrum. Thus,
our results do not conclusively suggest that a break does exist in the primordial spectrum,
rather they seem only to confirm the need for a spectral tilt.


\subsection{Power spectrum reconstruction} 
\begin{center}
\begin{figure}
\begin{center}
\psfrag{WMAP1}{\hspace{-2.5cm} Bridges06 WMAP1 analysis}
\psfrag{WMAP3}{\hspace{-0.2cm} dataset I}
\includegraphics[width=0.9\linewidth]{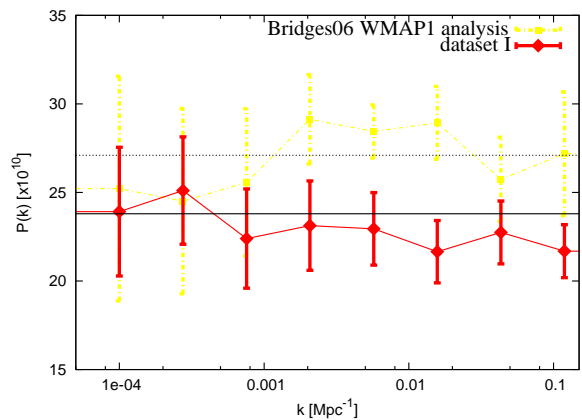}
\end{center}
  \caption{Reconstruction of the primordial power spectrum in 8 bands of $k$ (with $1\sigma$ errors) for dataset I (red) and our WMAP1 
  (Bridges06) analysis (yellow) 
  compared with the best fit 1-year (dotted) and 3-year (filled) H-Z spectrum.}
\label{figure6}
\end{figure}
\end{center}

\begin{center}
\begin{figure}
\begin{center}
\psfrag{WMAP3}{\hspace{-0.17cm} dataset I}
\psfrag{Lya}{\hspace{-0.7cm} dataset II}
\psfrag{LRG}{\hspace{-0.7cm} dataset III}
\includegraphics[width=0.67\linewidth, angle = -90]{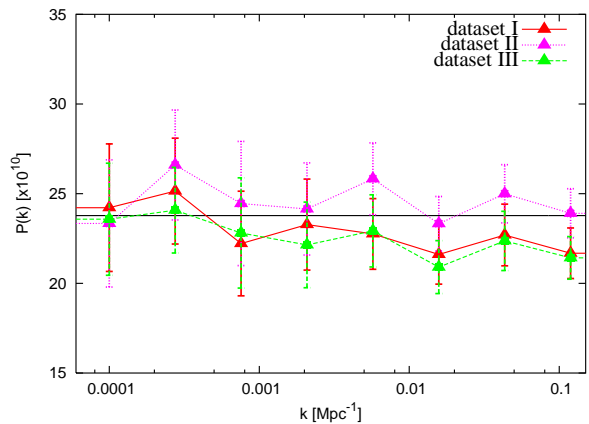}
\end{center}
  \caption{Reconstruction of the primordial power spectrum in 8 bands of $k$ (with $1\sigma$ errors) for dataset I (red), dataset II
   (magenta) and dataset III (green) 
  compared with the best fit (black) H-Z spectrum from dataset I.}
\label{figure7}
\end{figure}
\end{center}

Many previous attempts have been made to reconstruct the primordial spectrum directly
from data:
\citealt{Wang}; \citealt{Recon};  \citealt{Souradeep}; \citealt{SouradeepII};
\citealt*{Silk}; \citealt{Steen}; Bridges06. Most recently the WMAP team
\citep{SpergelII} attempted to reconstruct it using a set of amplitude bins in $k$ using WMAP3 data
alone. The method suffers from the natural side effect of imposing correlations 
between neighbouring bins and broadening other constraints. Therefore searching for
features by this method is unreliable and difficult. However from a model selection
viewpoint it does allow full freedom for the data to decide how many parameters are
required of the model. Our parameterisation linearly interpolates between eight
amplitude bins $a_n$ in $k$ on large scales between  0.0001 and 0.11 Mpc$^{-1}$
parameterised logarithmically with $k_{i+1} = 2.75k_i$ so that 
\begin{equation} 
P(k) = \left\{ \begin{array}{ll} \frac{(k_{i+1} - k)a_i + (k-k_i)a_{i+1}}{k_{i+1}-k_i},&
\mbox{$k_i <k < k_{i+1}$}\\ a_n,& \mbox{$k \geq k_n$}\end{array}\right. 
\end{equation} 
As expected an obvious tilt is discernible in our dataset I reconstruction (Fig. \ref{figure6})
the mean bin amplitudes deviating significantly from the best fit scale invariant spectrum
at high $k$,  not observable in our WMAP1 analysis. Though,  within 1$\sigma$ limits a H-Z spectrum can
still be fitted to both sets of data, however it would require a lower amplitude to
accommodate the values at large $k$. No cutoff is observed as was hinted at in our WMAP1
analysis; however uncertainty at this scale is dominated by high cosmic-variance, making
constraints inherently difficult. We find the addition of LRG data in dataset III provides little further
constraint  beyond that obtained from dataset I alone (see Fig. \ref{figure7}). The effect
of including Ly-$\alpha$ data in dataset II is however, marked. Firstly the overall amplitude is lifted, owing to
the larger $\sigma_8$ required by Ly-$\alpha$. Furthermore little obvious
tilt is discernible in full agreement with our analysis in Sec. \ref{h_z}. Encouragingly
similar features are seen in all three spectra such as the peaks at $0.003$ Mpc$^{-1}$ and
$0.05$ Mpc$^{-1}$. 

\subsection{Closed universe Inflation} 
\label{doran}
\begin{center}
\begin{figure}
\begin{center}
\includegraphics[width=0.9\linewidth]{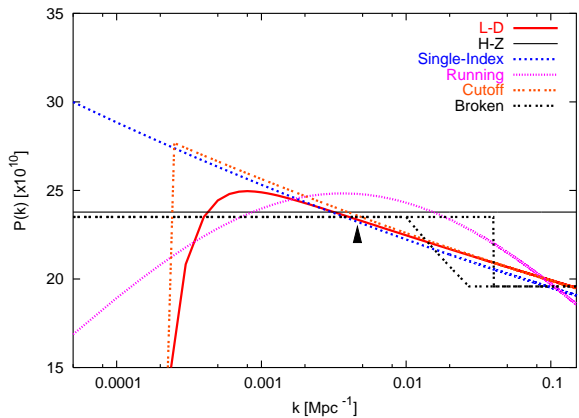}
\caption{L-D spectrum (solid-red) shown with the WMAP3 best fitting H-Z (solid-black), single-index with a cutoff
(dotted-orange), without (dotted-blue), a running index (dashed-pink) and broken (dotted-black).}
\label{figure8}
\end{center}
\end{figure}
\end{center}

\citet{Doran} arrived at a novel model spectrum by considering a boundary
condition that restricts the total conformal time available in the Universe, and
requires a closed geometry. The resultant predicted perturbation spectrum
encouragingly contains an exponential cutoff (as previously suggested
phenomenologically by \citealt{efstathiou}) at low $k$ that yields a
corresponding deficit in power in the CMB power spectrum. The shape of the
derived spectrum for a cosmology defined by $\Omega_0 = 1.04, \Omega_b h^2 =
0.0224, h = 0.6, \Omega_{cdm} h^2 = 0.110$ was parameterised by the function:
\begin{equation}
P(k) = A (1-0.023y)^2(1-\exp(-(y+0.93)/0.47))^2,
\end{equation}
where $y=\ln\left(\frac{k}{H_0/100}\times 3 \times 10^3\right)> -0.93$.
However we do know this form to be fairly stable to changes in cosmology, particularly in the position
of the cutoff. We therefore analyse this spectrum using dataset I, varying only
the amplitude within the best fit `concordance' cosmology found for the single
index model in Sec. \ref{h_z}. As expected the amplitude is reduced
when compared with our WMAP1 analysis ($A = 27.1 \pm 0.2$ Mpc$^{-1}$) due to the revised WMAP3
value for $\tau$.  At small scales the spectrum then lies roughly in line with a
tilted spectrum with $n_s \sim 0.96$ (see Fig. \ref{figure8}). For comparison the
plot also shows the other best fit spectra (including the bimodality in the
broken spectrum).

\subsection{Model Selection} We shall now turn to the fundamental inference. Which of
the models considered best describes the current data? As with our WMAP1 analysis we
perform the analysis in two stages to accommodate the L-D spectrum: the first, within
the full parameter space including all models bar L-D, while the second was carried
out within the fixed, best-fit single-index cosmology (as determined using dataset I) including L-D. While this
artificial division may seem flawed it should be remembered that within this
particular fixed cosmology the only spectrum that is favoured is that of the
single-index model itself, any other models should only be penalised. We chose two
sets of priors (see Table \ref{WMAP3_table1}) on $n_s$ and $n_{run}$, the wider of
which provides a very conservative range given the constraints available from current
data.  The results have been normalised to the single-index power law spectrum with
wide priors.
\begin{center}
\begin{table}
\begin{center}
\caption{Priors placed on the principal primordial spectral parameters.}
\vspace{1cm}
\begin{tabular}{|c||c|}
    \hline
 $A_s(\times 10^{-8})$     		& [15,55]\\
 $n_s$ (wide priors)		 	& [0.5,1.5]\\
 $n_s$  (narrow priors)		 		& [0.8,1.2]\\ 
 $n_{run}$ (wide priors)			& $[-0.15,0.15]$\\
 $n_{run}$ (narrow priors)				& $[-0.1,0.05]$\\
    \hline
\end{tabular}
\label{WMAP3_table1}
\end{center}
\end{table}
\end{center}

\subsubsection{Full cosmological parameter space exploration}
\label{full}
Our corresponding analysis of WMAP1 data was unable to make any
conclusive model selection; with statistical uncertainty dominating results. With
WMAP3 data and the method of nested sampling however we have arrived at a cusp in cosmological 
model selection. With dataset I, Table \ref{WMAP3_table2}
confirms the disfavouring of a scale-invariant spectrum that we saw in the parameter constraints in Sec.
\ref{h_z}, though not at a decisive level according
to the Jeffrey's scale. As expected from the parameter constraints
a running spectrum is preferred, but only at a significant level if narrow priors are
chosen on $n_s$ and $n_{run}$. Large scale power is suppressed in WMAP3 as it was
in WMAP1, with little improvement in uncertainty: both datasets are almost
cosmic-variance limited on these scales so a cutoff is
preferred by a significant margin of over 2 units in log-evidence. The
evidence in favour of a broken spectrum is likely only due to its mimicking of a
tilted spectrum as opposed to modelling of an abrupt break.

Inclusion of Ly-$\alpha$ in dataset II confirms the conclusions drawn in Section \ref{h_z},
showing a significantly reduced evidence in favour of spectral running (with narrow priors),
from a +1.2 with dataset I  to -1.3, a significant evidence difference between the two
datasets.    Unsurprisingly all three data combinations exhibit significant  evidences in
favour of a spectral cutoff, due presumably in each case to the decrement in WMAP3 data 
alone. 

\begin{center} 
\begin{table*} 
\begin{center} 
\caption{Differences of log-evidences with respect to single-index model (with the widest
priors) using the full
cosmological parameter space. [Note evidence comparisons can only be made between models
using the same dataset.]}
\begin{tabular}{|c||c||c||c|}
    \hline
 \textbf{Model} & datset I & + Ly-$\alpha$ & + LRG\\
    \hline
 H-Z     				& $-0.7$ $\pm$ 0.3& $-0.1$$\pm$ 0.3& $-0.4$ $\pm$ 0.3\\
 Single-Index (wide priors)		& +0.0 $\pm$ 0.3& +0.0$\pm$ 0.3& +0.0 $\pm$ 0.3\\
 Single-Index (narrow priors)		 		& +1.0 $\pm$ 0.3& +0.2$\pm$ 0.3& +0.7 $\pm$ 0.3\\ 
 Running (wide priors)			& $-2.9$ $\pm$ 0.3& $-1.6$$\pm$ 0.3& $-1.8$ $\pm$ 0.3\\
 Running (narrow priors on $n_s$)		& +0.4 $\pm$ 0.3& $-0.7$$\pm$0.3& +1.7 $\pm$ 0.3\\
 Running (narrow priors on $n_{run}$ \& $n_s$)	& +1.2 $\pm$ 0.3& $-1.3$$\pm$ 0.3& +1.0 $\pm$ 0.3\\
 Cutoff 				& +2.3 $\pm$ 0.3& +1.7$\pm$ 0.3& +2.9 $\pm$ 0.3\\
 Barriga				& +1.0 $\pm$ 0.3& +1.2$\pm$ 0.3& +0.9 $\pm$ 0.3\\
    \hline
\end{tabular}
\label{WMAP3_table2}
\end{center}
\end{table*}
\end{center}

\subsubsection{Primordial parameter space exploration} The large preference in
favour of a spectral cutoff and increasingly tight constraints on the universal
geometry being marginally closed (a fact that is particularly reinforced when
examining the LRG data \citet{TegmarkII}), suggests that the L-D spectrum may
provide a very good fit to current data. For the remainder of this analysis we will
examine the L-D spectrum using dataset I only.

As with the WMAP1 analysis, dataset I, also prefers the L-D
spectrum, now by a slightly larger log-evidence (see Table \ref{WMAP3_table3}.), which according
to Jeffreys' scale now constitutes a decisive model selection. In Bridges06 we concluded
that the `significant' model detection was due to the form of the spectrum on large scales
i.e. its exponential cutoff. However on small scales it behaves much like a tilted spectrum
with slight running, a form we now know fits the data very well. Could it be that these
small scale features are driving this model selection?

To test this hypothesis we have analysed three \emph{hybrid} spectra, where we have divided the L-D
 spectrum about $k\approx 0.008$ Mpc$^{-1}$ (denoted by the arrow in Fig. \ref{figure8}) which corresponds 
 loosely with the angular scale
($l=8$) where a cutoff ceases to be observed. We have then spliced both sections, about this
point, with various single-index spectra at large or small scales. Model A combines a single-index
 spectrum below $k = 0.0008$ Mpc$^{-1}$ with L-D thereafter; Model B: an exponential
cutoff from L-D with a fixed single-index model ($n= 0.94$) for $k>0.0008$ and Model C: as for
B but with a \emph{varying} single-index model. 

The results (see Table \ref{WMAP3_table3}) bear out our assertion; models A and B have roughly
identical log-evidence values to the original L-D, demonstrating the data to be essentially indifferent to the presence of a
cutoff. In comparison, model C is typically just as good as a power law spectrum (i.e. an evidence close to 0). This suggests
that the L-D spectrum is attractive to the data as 
it naturally incorporates a tilt without the need to parameterise it. However the tilt present in the L-D spectrum ($n_s \sim 0.96$) 
does coincide well with the best fit values obtained in Sec. \ref{h_z} for a power law spectrum (which are fairly invariant among the 
datasets I, II or III). This fact coupled with a significant evidence preference for a large scale spectral cutoff suggests 
the L-D spectrum does provide a uniquely good fit to current data. 
\begin{center}
\begin{table}
\begin{center}
\caption{Differences of log-evidences (primordial parameters only) for all models with respect to the single-index models preferred
cosmology: $\Omega_0 = 1.035, \Omega_b h^2 = 0.0221, h = 0.58, \Omega_{cdm} h^2 = 0.112$, with our previous
WMAP1 results for comparison.}
\vspace{1cm}
\begin{tabular}{|c||c||c|}
    \hline
\textbf{Model} & WMAP1 (Bridges06)  & dataset I\\
    \hline
 Single-Index & +0.0 $\pm$ 0.5 & +0.0 $\pm$ 0.2\\
 L-D  & +4.1 $\pm$ 0.5 & +5.2 $\pm$ 0.2\\
 L-D (A) & -- & +5.0 $\pm$ 0.2\\
 L-D (B) & -- & +5.2 $\pm$ 0.2\\
 L-D (C) & -- & +0.9 $\pm$ 0.2\\
    \hline
\end{tabular}
\label{WMAP3_table3}
\end{center}
\end{table}
\end{center}

\section{Conclusions} 
A scale-invariant spectrum is now largely disfavoured by the dataset I
with a spectral index $n_s=0.95 \pm 0.02$ deviating by at least 2$\sigma$ from
$n_s=0$. Moreover a running spectrum ($n_{run} = -0.038 \pm 0.030$) is now significantly
 preferred but only using
the most constraining prior. 
The addition of Ly-$\alpha$ forest data improves all constraints but does not
alter the preferred spectral tilt greatly. It does however, along with LRG data, suggest a
significantly smaller running index ($n_{run} = -0.015 \pm 0.015$, $n_{run} = 0.01 \pm 0.05$).
This tension has previously been analysed by \citet{Seljak} who conclude that such discrepancies, even 
though at the 2$\sigma$ level are consistent with normal statistical fluctuations between datasets.   
A power law spectrum with a cutoff provides the best evidence fit in our full parameter space study
with a significant evidence ratio of roughly 2 units across all three datasets. The similarity of this
cutoff model with the L-D spectrum suggests the latter should also provide a very good fit. This is
indeed borne out with decisively large evidence ratios within our limited
primordial-only analysis.

\section*{Acknowledgements}
This work was conducted in cooperation with SGI/Intel
utilising the Altix 3700 supercomputer (UK National Cosmology Supercomputer) at DAMTP Cambridge
supported by HEFCE and PPARC.
We thank S. Rankin and V. Treviso for their computational assistance. 
MB was supported by a Benefactors Scholarship at St. John's College, Cambridge and an Isaac
Newton Studentship.

\appendix

\label{lastpage}


\begin{thebibliography}{99}


\bibitem[\protect\citeauthoryear{Abazajian et al.}{2003}]{sloan}
Abazajian K., et al., 2003, Astrophys. J., 126, 2081
\bibitem[\protect\citeauthoryear{Adams, Ross \& Sarkar}{Adams et al.}{1997}]{Adams}
Adams J.A., Ross G.G. \& Sarkar S., 1997, Nucl. Phys. B. 503, 405
\bibitem[\protect\citeauthoryear{Barriga et al.}{2001}]{barriga}
Barriga J., Gaztanaga E., Santos M.G., Sarkar S., 2001, MNRAS, 324, 977 
\bibitem[\protect\citeauthoryear{Basset et al.}{2004}]{Basset}
Basset B.A., Corasaniti P.S., Kunz, M., Astrophys. J. Lett., 2004, 617, L1
\bibitem[\protect\citeauthoryear{Beltran et al.}{2005}]{Beltran}
Beltran M., Garcia-Bellido J., Lesgourgues J., Liddle A., Slosar A., 2005, 
Phys. Rev. D, 71, 063532 
\bibitem[\protect\citeauthoryear{Bennett et al.}{2003}]{WMAP1}
Bennett C.L., et al., 2003, Astrophys. J. Suppl., 148, 1
\bibitem[\protect\citeauthoryear{Bridges et al.}{2006}]{Bridges}
Bridges M., Lasenby A.N., Hobson, M.P., 2006, MNRAS, 369, 1123
\bibitem[\protect\citeauthoryear{Bridle et al.}{2003}]{Recon}
Bridle S., Lewis A., Weller J., Efstathiou G., 2003, MNRAS, 342, L72
\bibitem[\protect\citeauthoryear{Contaldi et al.}{2003}]{Contaldi}
Contaldi C.R., Peloso M., Kofman L., Linde A., 2003, J. Cosmol. Astropart. Phys., 7, 2
\bibitem[\protect\citeauthoryear{Dickinson et al.}{2004}]{VSA}
Dickinson C. et al., 2004, MNRAS, 353, 732
\bibitem[\protect\citeauthoryear{Drell et al.}{2000}]{Drell}
Drell P.S., Loredo T.J., Wasserman, I., 2000, Astrophys. J., 530, 593
\bibitem[\protect\citeauthoryear{Efstathiou}{2003}]{efstathiou}
Efstathiou G., 2003, MNRAS, 346, 26 
\bibitem[\protect\citeauthoryear{Freedman et al.}{2001}]{HST}
Freedman W.L., et al., 2001, Astrophys. J., 553, 47
\bibitem[\protect\citeauthoryear{Hannestad}{2004}]{Steen}
Hannestad S., 2004, J. Cosmol. Astropart. Phys., 4, 2
\bibitem[\protect\citeauthoryear{Hinshaw et al.}{2007}]{WMAP3}
Hinshaw G. et.al., 2007, Astrophys. J. Suppl., 170, 288
\bibitem[\protect\citeauthoryear{Hobson et al.}{2003}]{Bridle}
Hobson M.P., Bridle S.L., Lahav O., 2002, MNRAS, 335, 377
\bibitem[\protect\citeauthoryear{Hobson \& McLachlan}{2003}]{McLachlan}
Hobson M.P., McLachlan C., 2003, MNRAS, 338, 765
\bibitem[\protect\citeauthoryear{Jaffe}{1996}]{Jaffe}
Jaffe A., 1996, Astrophys. J., 471, 24
\bibitem[\protect\citeauthoryear{Jeffreys}{1961}]{Jeffreys}
Jeffreys H., 1961, \emph{Theory of Probability}, 3rd ed., Oxford University Press
\bibitem[\protect\citeauthoryear{John \& Narlikar}{2002}]{John}
John M.V., Narlikar J.V., 2002, Phys. Rev. D, 65, 043506
\bibitem[\protect\citeauthoryear{Kuo et al.}{2004}]{ACBAR}
Kuo C.L. et al., 2004, Ap. J., 600, 32
\bibitem[\protect\citeauthoryear{Lasenby \& Doran}{2005}]{Doran}
Lasenby A.N., Doran, C., 2005, Phys.Rev. D 71, 063502
\bibitem[\protect\citeauthoryear{Lewis \& Bridle}{2002}]{cosmomc}
Lewis A. and Bridle S., 2002, Phys. Rev. D, 66, 103511
\bibitem[\protect\citeauthoryear{Marshall et al.}{2003}]{Marshall}
Marshall P.J., Hobson M.P., Slosar A., 2003, MNRAS, 346, 489

\bibitem[\protect\citeauthoryear{McDonald et al.}{2006}]{McDonaldI}
McDonald P., et al., 2006, Astrophys. J. Suppl., 163, 80-109
\bibitem[\protect\citeauthoryear{McDonald et al.}{2005}]{McDonaldII}
McDonald P., et al., 2005, Astrophys. J., 635, 761-783

\bibitem[\protect\citeauthoryear{Mukherjee, Parkinson \& Liddle}{Mukherjee et al.}{2006}]{Mukherjee}
Mukherjee, P. Parkinson D. Liddle, A., 2006, Astrophys. J., 638, L51-L54
\bibitem[\protect\citeauthoryear{Mukherjee and Wang}{2003}]{Wang}
Mukherjee P. \& Wang Y., 2003, Astrophys. J., 593, 38
\bibitem[\protect\citeauthoryear{Niarchou et al.}{2004}]{Niarchou}
Niarchou A., Jaffe A., Pogosian L., 2004, Phys.Rev. D 69 063515
\bibitem[\protect\citeauthoryear{Percival et al.}{2001}]{2dF}
Percival W.J. et al., 2001, MNRAS, 327, 1297
\bibitem[\protect\citeauthoryear{Readhead et al.}{2004}]{CBI}
Readhead A.C.S. et al., 2004, Astrophys. J.,, 609, 498--512
\bibitem[\protect\citeauthoryear{Parkinson et al.}{2006}]{Parkinson}
Parkinson D., Mukherjee P., Liddle A.R., 2006, Phys. Rev. D., 73, 123523
\bibitem[\protect\citeauthoryear{Saini et al.}{2004}]{Saini}
Saini T.D., Weller J., Bridle S.L., 2004, MNRAS, 348, 603

\bibitem[\protect\citeauthoryear{Seljak et al.}{2006}]{Seljak}
Seljak U., Slosar A., McDonald P., 2006, J. Cosmol. Astropart. Phys., 10, 14

\bibitem[\protect\citeauthoryear{Skilling}{2004}]{Skilling}
Skilling J., 2004, Bayesian Inference and Maximum Entropy Methods in Science and Engineering, Ed. R. Fisher et al., American Inst.
Phys. conf. proc., 735, 395
\bibitem[\protect\citeauthoryear{Slosar et al.}{2003}]{Slosar}
Slosar A. et al., 2003, MNRAS, 341, L29
\bibitem[\protect\citeauthoryear{Shafieloo \& Souradeep}{2004}]{Souradeep}
Shafieloo, A., Souradeep, T., 2004, Phys. Rev. D, 70, 043523 

\bibitem[\protect\citeauthoryear{Shaw et al.}{2007}]{Shaw}
Shaw J.R., Bridges M., Hobson M.P., 2007, MNRAS, 378, 1365-1370
\bibitem[\protect\citeauthoryear{Sinha \& Souradeep}{2006}]{SouradeepII}
Sinha, R., Souradeep, T., 2006, Phys. Rev. D., 74, 043518
\bibitem[\protect\citeauthoryear{Spergel et al.}{2003}]{Spergel}
Spergel D.N. et al., 2003, Astrophys. J. Suppl., 148, 175
\bibitem[\protect\citeauthoryear{Spergel et al.}{2007}]{SpergelII}
Spergel D.N. et al., 2007, Astrophys. J. Suppl., 170, 377


\bibitem[\protect\citeauthoryear{Tegmark et al.}{2006}]{TegmarkII}
Tegmark M. et al., 2006,Phys. Rev. D, 74, 123507 

\bibitem[\protect\citeauthoryear{Tocchini-Valentini, Douspis \& Silk}{Tocchini-Valentini et al.}{2005}]{Silk}
Tocchini-Valentini D., Douspis M., Silk J., 2005, MNRAS, 359, 31
\bibitem[\protect\citeauthoryear{Trotta}{2007}]{Trotta}
Trotta R., 2007, MNRAS, 378, 72
\bibitem[\protect\citeauthoryear{Viel, Haehnelt \& Lewis}{2006}]{Viel}
Viel M., Haehnelt M.G., Lewis A., 2006, MNRAS, 370, L51


\bibitem[\protect\citeauthoryear{Wang}{1994}]{Wang}
Wang Y., 1994, Phys. Rev. D, 50, 6135
\end{thebibliography}
\end{document}